\documentclass[reprint,amsmath,amssymb,aps,pra]{revtex4-2}

\usepackage{graphicx,wrapfig,lipsum}
\usepackage{subeqnarray}
\usepackage{url,hyperref} 
\usepackage{color}
\usepackage{ulem}
\usepackage{verbatim}
\usepackage{mathtools}

\usepackage{natbib}
\usepackage{float}

\newcommand{\wt}[1]{\widetilde{#1}}
\newcommand{\eq}[1]{#1^{(\mathrm{eq})}}

\newcommand{\req}[1]{Eq.\,({\ref{#1}})}
\newcommand{\reqs}[2]{Eqs.\,({\ref{#1}}-{\ref{#2}})}
\newcommand{\mk}{|\boldsymbol{k}|}
\newcommand{\ft}[1]{\widetilde{\boldsymbol{#1}}}
\newcommand{\hatv}[1]{\hat{\boldsymbol{#1}}}

\begin{document}
\title{Dynamic Magnetic Response of Quark-Gluon Plasma to Electromagnetic Fields}

\author{Christopher Grayson}
\email{chrisgray1044@email.arizona.edu}

\author{Martin Formanek}
\email{martinformanek@email.arizona.edu}

\altaffiliation[Address after May 1, 2022: ]{Institute of Physics AS CR, ELI-Beamlines project, Za Radnicí 835, 25241 Dolní Břežany, Czech Republic.}

\author{Johann Rafelski}
\email{johannr@arizona.edu}
\affiliation{Department of Physics,
The University of Arizona,
Tucson, AZ, 85721, USA}
\author{Berndt M\"uller}
\email{mueller@phy.duke.edu}
\affiliation{Department of Physics,
Duke University,
Durham, NC 27708-0305, USA}

\date{4/29/2022}
\begin{abstract}
We investigate the electromagnetic response of a viscous quark-gluon plasma in the framework of the relativistic Boltzmann equation with current conserving collision term. Our formalism incorporates dissipative effects at all orders in linear response to the electromagnetic field while accounting for the full space- and time-dependence of the perturbing fields. As an example, we consider the collision of two nuclei in a stationary, homogeneous quark-gluon plasma. We show that for large collision energies the induced magnetic fields are governed by the response of quark-gluon plasma along the light-cone.  In this limit we derive an analytic expression for the magnetic field along the beam axis between the receding nuclei and show that its strength varies only weakly with collision energy for $\sqrt{s_\mathrm{NN}} \ge 30$\,GeV. 
\end{abstract}
\maketitle

%%%%%%%%%%%%%%%%%%%%%%%%%%%%%%%%%%%%%%%%%%%%%%%%%%%%%%%%%%%%%%%%%%%%%%%

\section{Introduction}

The electromagnetic fields generated by colliding relativistic heavy ions at RHIC and LHC are some of the largest in nature, on the order of $m_\pi^2/e \approx 10^{14}$\,T, but they are very short-lived with a time constant $t_{\text{coll}}= 2 R/\gamma \sim 10^{-25}\,\textrm{s}$, where $R$ is the nuclear radius and $\gamma$ is the Lorentz factor. These fields occur in the presence of a quark-gluon plasma (QGP), the high temperature matter that forms in the region of overlap between the colliding ions. Within the plasma these strong electromagnetic fields generate a response that scales with $Z\alpha$, where $Z$ is the charge of the ions and $\alpha$ is the fine-structure constant. The polarization current is subject to the strong dissipative effects in the QGP that render the QGP an over-damped system which sustains slowly decaying magnetic wake fields.

In the past, most analytic calculations \cite{Tuchin:2010vs,Deng:2012pc,McLerran:2013hla,Tuchin:2013apa,Gursoy:2014aka,Li:2016tel,Roy:2015kma} assumed that the response of the QGP is static and obtained the late time dependence of the induced magnetic field by solving Maxwell's equations in the presence of static electric conductivity in a  hydrodynamically evolving QGP. These approaches ignore the fact that the time scale of variation of the electromagnetic fields, $t_{\text{coll}}$, is shorter than the characteristic time scales of the plasma response, given by the plasma frequency $\omega_p$ and the collisional damping rate $\kappa$, as shown in Table~\ref{tab:timescales}. Some numerical calculations \cite{Inghirami:2016iru, Inghirami:2019mkc} have attempted to incorporate the dynamical response of QGP by numerically solving the coupled magnet-hydrodynamic equations for a conducting quark-gluon plasma in the presence of the colliding nuclear charges. More recent calculations \cite{Yan:2021zjc,Wang:2021oqq} also incorporate the frequency and wave-vector dependence of QGP response to electromagnetic fields by solving the coupled Vlasov-Boltzmann--Maxwell equations. The disadvantage of these approaches, apart from the fact that they are prone to numerical imprecision, is that they do not easily allow for parametric predictions of how the medium response depends on the properties of the QGP and energy of the colliding nuclei.

Here we address this issue by calculating the response of a QGP to heavy ion fields using the full frequency and wave-vector dependent relativistic polarization tensor previously obtained in \cite{Formanek:2021blc} by using relativistic kinetic theory to derive a gauge invariant polarization tensor within the linear response formalism. As we are interested in high collision energies, we consider the case of a baryon number-symmetric plasma. The result of \cite{Formanek:2021blc} accounts for both, collective plasma response and viscous effects, and allows us to obtain analytical expressions for the induced fields and currents that depend parametrically on the electromagnetic Debye mass $m_D$ and the average collision rate $\kappa$. We expect our results to have validity as an effective description of the plasma response that is independent of the detailed assumptions for the QGP transport model.

The average collision rate $\kappa$ is inversely related to the mean-free time of partons in the plasma, $\kappa =1/\tau_{\text{rel}}$, also called relaxation time. We will show in Sect.~\ref{sec:energymomcons} that, for a particle-antiparticle plasma, current conservation implies energy momentum conservation to first order in the plasma response. The collision term generates damping in the evolution of the one-particle distribution function $f(x,p)$ in response of the electromagnetic currents and velocity gradients. This gives the plasma dissipative properties parametrized by the shear viscosity and the electric conductivity, which are related to the collision rate $\kappa$ by (see e.~g.\ \cite{Florkowski:2017olj}):
\begin{equation}
    \eta = \frac{s T}{ 5 \kappa}\,, \qquad \sigma_0 = \frac{e^2T^2}{3 \kappa}\,,
\end{equation}
where $s$ denotes the entropy density and $T$ is the temperature. The treatment of conservation laws in transport coefficients in the relaxation time approximation is discussed in \cite{Rocha:2021zcw}, where it is shown that in the case of electrical conductivity and shear viscosity the BGK (Bhatnagar-Gross-Krook) term \cite{Bhatnagar:1954zz} is the only required correction to the collision term.

Here we focus on the damping effect of the collision rate $\kappa$ on the induced magnetic fields in nuclear collisions. The collision term generates a nonvanishing conductivity via the imaginary part of the polarization tensor. This conductivity manifests itself in poles of the resummed propagator in the lower complex $\omega$-plane that generate long-range tails or wake fields that extend far beyond the collision time. In Table~\ref{tab:timescales} we collect the relevant timescales of the problem in ascending order.  The collision time $t_{\text{coll}}$ is much shorter than all other relevant time scales. The collective oscillations of the plasma are highly damped by the large value of the relaxation time giving rise to over-damped behavior.

\begin{table}[h!]
\caption{\label{tab:timescales}
Approximate time scales relevant to the electromagnetic response of QGP for an Au+Au collisions at $\sqrt{s_{\text{NN}}}=200\,$MeV with QGP temperature $T=300$\,MeV. Time scales are shown in ascending order.}
\begin{ruledtabular}
\begin{tabular}{ccc}
\textrm{Time Scale}&
\textrm{Formula}&
\multicolumn{1}{c}{\textrm{Time (fm/c)}}\\
\colrule
 \textrm{Collision Time} & $t_\text{coll} = 2 R/\gamma$ & 0.086\footnote{Calculated using the Gaussian radius $R = 4.33$\,fm defined in \req{eq:radius}.} \\
\textrm{Relaxation Time} & $\tau_\text{rel} = 1/\kappa$ & 0.36 \\

\textrm{Freeze-out time} & $t_f$ & $5$\footnote{Estimated using 2+1 dimensional hydrodynamic evolution \cite{Song:2007ux}.}\\
\textrm{ Decay Time} & $t_{\sigma} = 1/\sigma_0 = \kappa/\omega_p^2$ & 59\footnote{The decay time is the large damping $\kappa/\omega_p$ expansion of the plasma oscillation frequency \req{eq:plasmafreq}. } \\
\end{tabular}
\end{ruledtabular}
\end{table}

The layout of this work is as follows. We begin by solving Maxwell's equations in a homogeneous, stationary, polarizable medium in the limit of linear response in Sect.~\ref{sec:Maxwell2}. We go on to summarize the calculation of the electromagnetic polarization tensor of QGP following \cite{Formanek:2021blc} and briefly discuss energy-momentum conservation in Sect.~\ref{sec:linresp}, where we also present our choice of the QGP parameters. We then explore the magnetic fields generated in a nuclear collision in Sect.~\ref{sec:results}. We compare the magnetic field calculated with full space-time dependence in the polarization tensor to previously used approximations and find an analytic expression for the late time magnetic field at the collision center. We summarize our results in Sect.~\ref{sec:Conclusions}.

%%%%%%%%%%%%%%%%%%%%%%%%%%%%%%%%%%%%%%%%%%%%%%%%%%%%%%%%%%%%%%%%%%%%%%%%%%%%%%%%%%%%%%%%%%%%%%%%

\section{Maxwell equations in QGP}\label{sec:Maxwell2}

We begin by solving Maxwell's equations in an infinite homogeneous and stationary polarizable medium. In this medium Maxwell's equations take on the usual form
\begin{equation}
\partial^{[\mu}F^{\nu \rho]}(x) =0, \quad \partial_{\mu}F^{\mu \nu}(x) = \mu_0 J^{\nu}(x)\,,
\end{equation}
or in Fourier space using the replacement $\partial^{\mu} \rightarrow -i k^{\mu} $
\begin{equation}
-i k^{[\mu}\wt{F}^{\nu \rho]}(k) =0, \quad -i k_{\mu}\wt{F}^{\mu \nu}(k) = \mu_0 \wt{J}^{\nu}(k)\,.
\end{equation}
The properties of the medium are introduced by writing the 4-current $\wt{J}^{\mu}$ in terms of its induced and external parts
\begin{equation}
 \wt{J}^{\mu}(k) = \wt{j}_{\mathrm{ext}}^{\mu}(k)+ \wt{j}_{\mathrm{ind}}^{\mu}(k)\,.
\end{equation}
The induced current $\wt{j}_{\mathrm{ind}}^{\mu}$ to leading order is given by the polarization tensor through the covariant formulation of Ohm's law \cite{Starke:2014tfa}
\begin{equation}\label{eq:linresp2}
    \wt{j}_{\mathrm{ind}}^{\mu}(k) = {\Pi^{\mu}}_{\nu}(k) \wt{A}^{\nu}(k)\,,
\end{equation}
where $\Pi^{\mu}_{\nu}$ is the polarization tensor and $\wt{A}^{\mu}$ is the 4-vector potential. The electromagnetic tensor $\wt{F}^{\mu \nu}$ can be written in terms of the 4-vector potential
\begin{equation} \label{eq:emten}
\wt{F}^{\mu \nu}(k) =  - i \left(k^{\mu} \wt{A}^{\nu}(k) - k^{\nu} \wt{A}^{\mu}(k)\right)\,.
\end{equation}
Solving Maxwell's equations in the Lorentz gauge $k \cdot \wt{A}=0$ one finds the usual expression
\begin{equation}\label{eq:Amu}
\wt{A}^{\mu}(k)= -\frac{\mu_0}{k^2}\left(\wt{j}_{\mathrm{ext}}^{\mu}(k)+ \wt{j}_{\mathrm{ind}}^{\mu}(k)\right)\,
\end{equation}
where $\mu_0$ denotes the magnetic permittivity of the vacuum.

We proceed by algebraically solving for the self-consistent potential that contains the back-reaction of the induced current on the field to all orders in the Debye screening mass $m_D$. To do this, we first note that in a homogeneous medium the response depends only on two independent scalar polarization functions $\Pi_\parallel$ and $\Pi_\perp$ describing polarization in the parallel and transverse directions relative to the wave-vector $\boldsymbol{k}$ \cite{Weldon:1982aq}. The polarization tensor may be written in terms of these polarization functions as
\begin{equation}\label{eq:poltensgen}
 \Pi^{\mu \nu}(k,u) = \Pi_\parallel(k) L^{\mu \nu}(k,u) + \Pi_\perp(k) S^{\mu \nu}(k,u)\,,
\end{equation}
where $k^\mu$ is the 4-momentum of the field and $u^\mu$ is the 4-velocity of the medium. Conventions for the longitudinal and transverse projection tensors, $L^{\mu \nu}$ and  $S^{\mu \nu}$ respectively, may be found in \cite{Melrose:2008}. These tensors are reproduced here for convenience
\begin{multline}
     L^{\mu \nu} \equiv \frac{k^2}{(k\cdot u)^2-k^2}\bigg[ \frac{ k^{\mu}u^{\nu}}{(k\cdot u)}+ \frac{ k^{\nu}u^{\mu}}{(k\cdot u)}\\ -\frac{k^2u^{\mu}u^{\nu}}{(k\cdot u)^2}  -\frac{k^{\mu}k^{\nu}}{k^2} \bigg]\,,
\end{multline}
\begin{multline}
     S^{\mu \nu} \equiv g^{\mu \nu} +\frac{1}{(k\cdot u)^2-k^2}\bigg[ k^{\mu}k^{\nu} \\
     -(k\cdot u)( k^{\mu}u^{\nu}+k^{\nu}u^{\mu})+k^2u^{\mu}u^{\nu}\bigg]\,.
\end{multline}
These projections are equivalent to ones defined in \cite{Weldon:1982aq} up to an overall normalization. To simplify the calculation the wave-vector $\boldsymbol{k}$ is chosen, without loss of generality, to point along the third spatial direction ($\mu=3$):
 \begin{equation}\label{eq:poltenmat}
    \Pi^{\mu}_{\nu}(\omega,\boldsymbol{k}) = \left[
    \begin{array}{cccc}
-\frac{|\mathbf{k}|^2}{\omega^2}\Pi_{\parallel}& 0 & 0 & \frac{|\mathbf{k}|}{\omega}\Pi_{\parallel} \\
 0 & \Pi_{\perp} & 0 & 0 \\
 0 & 0 & \Pi_{\perp} & 0 \\
 -\frac{|\mathbf{k}|}{\omega}\Pi_{\parallel} & 0 & 0 & \Pi_{\parallel} \\ 
\end{array}
\right]\,.
\end{equation}
Utilizing this decomposition the spatial component of the potential $\ft{A}$ is expressed as
\begin{equation}
\ft{A} = \wt{A}_\parallel \hatv{k} + \ft{A}_\perp\,,
\end{equation}
which implies
\begin{equation}
 \wt{A}_{\parallel} = \frac{\boldsymbol{k} \cdot  \wt{\boldsymbol{A}}}{|\boldsymbol{k}|}, \quad   \wt{\boldsymbol{A}}_{\perp} = \wt{\boldsymbol{A}} -  \wt{A}_{\parallel}\hat{\boldsymbol{k}}\,,
\end{equation}
with analogous definitions for the current, $\wt{j}_{\parallel}$ and $\wt{j}_{\perp}$. Note that the Lorentz gauge condition $\partial_\mu A^\mu = 0$ implies
\begin{equation}\label{eq:apar}
\wt{A}_\parallel = \frac{\omega}{ |\boldsymbol{k}|}\wt{\phi}\,
\end{equation} 
with $\phi=A^0$. The resulting induced charge can be calculated using the projected polarization tensor \req{eq:poltenmat}:
\begin{equation}
    \wt{\rho}_\text{ind}(\omega,\boldsymbol{k})  = \Pi^0_\nu \wt{A}^\nu = -\frac{|\boldsymbol{k}|^2}{\omega^2} \Pi_{\parallel}\wt{\phi} +  \frac{|\boldsymbol{k}|}{\omega}\Pi_{\parallel} \wt{A}_{\parallel}\,.
\end{equation}
For the Lorentz gauge condition \req{eq:apar} one finds
\begin{equation}\label{eq:indch}
    \wt{\rho}_\text{ind}(\omega,\boldsymbol{k})  = \Pi_{\parallel}\wt{\phi} \left(1 -\frac{|\boldsymbol{k}|^2}{\omega^2}\right)\,.
\end{equation}
Similarly,
\begin{equation}\label{eq:indjpar}
\wt{j}_{\parallel\text{ind}}(\omega,\boldsymbol{k})  =  \Pi^z_\nu \wt{A}^\nu  = \Pi_{\parallel}  \frac{\omega}{\mk}\wt{\phi}\left(1-\frac{|\boldsymbol{k}|^2}{\omega^2} \right)\,,
\end{equation}
expressing current conservation, $\partial_\mu j^\mu = 0$. The induced transverse current is
\begin{equation}\label{eq:indjperp}
   \boldsymbol{j}_{\perp\text{ind}}(\omega,\boldsymbol{k})  =  \Pi_{\perp} \wt{A}_\perp\,.
\end{equation}

Solving for the potential on both sides of \req{eq:Amu} with the help of \reqs{eq:indch}{eq:indjperp} gives the self-consistent solutions,
\begin{align}\label{eq:phi}
&\wt{\phi}(\omega,\boldsymbol{k}) = \frac{\wt{\rho}_\text{ext}(\omega,\boldsymbol{k})}{\varepsilon_0(\boldsymbol{k}^2-\omega^2) \left(\Pi_{\parallel}/( \omega^2\varepsilon_0)+1\right) }\,, \\\label{eq:aperp}
&\ft{A}_\perp(\omega,\boldsymbol{k}) = \frac{\mu_0 \ft{j}_{\perp \text{ext}}(\omega,\boldsymbol{k})}{\boldsymbol{k}^2 - \omega^2 - \mu_0 \Pi_{\perp}}\,.
\end{align}
The gauge condition \req{eq:apar} gives the self-consistent potential $\wt{A}_\parallel$. These self-consistent potentials determine the electric and magnetic fields via the usual relations
\begin{equation}\label{eq:ftfields}
\ft{B}(\omega,\boldsymbol{k}) = i\boldsymbol{k} \times \ft{A}_\perp\,, \quad \ft{E}(\omega,\boldsymbol{k}) = -i \boldsymbol{k} \wt{\phi} + i \omega \ft{A}\,.
\end{equation}

The magnetic field of interest to us is obtained by Fourier transforming these expressions back to position space (for details see Appendix \ref{sec:magf}). We note here that it is important to calculate the fields with the resummed propagator in \reqs{eq:phi}{eq:aperp}, rather than by using its series representation, in order to correctly capture the pole structure of the propagator that governs the late-time dependence of the magnetic field. For ease of calculation, we specify the external 4-current using two colliding Gaussians charge distributions normalized to the nuclear rms radius $R$ and charge $Z$:
\begin{multline}\label{eq:rhoext}
\rho_{\text{ext}\pm }(t,\boldsymbol{x}) = \frac{Zq\gamma}{\pi^{3/2}R^3}e^{-\frac{1}{R^2}(x\mp b/2)^2}e^{-\frac{1}{R^2}y^2}\\
\times e^{-\frac{\gamma^2}{R^2}(z\mp \beta t)^2}\,,
\end{multline}
where $\gamma$ and $\beta$ are the Lorentz factor and speed, respectively, and $b$ is the impact parameter of the collision. The plus and minus signs indicate motion in the $\pm \hat{z}$-direction (beam-axis). This charge distribution corresponds to the vector current
\begin{equation}\label{eq:jext}
\boldsymbol{j}_{\text{ext}\pm}(t, \boldsymbol{x}) = \pm \beta \hatv{z} \rho_{\text{ext}\pm}(t, \boldsymbol{x})\,.
\end{equation}
Further details of the external charge distribution for two colliding nuclei are presented in Appendix \ref{sec:freechg}..

%%%%%%%%%%%%%%%%%%%%%%%%%%%%%%%%%%%%%%%%%%%%%%%%%%%%%%%%%%%%%%%%%%%%%%%%%%%%%%%%%%%%%%%%%%%%%%%%%%%%%%%%%%%%%%%%%%%%

\section{The QGP Polarization tensor}\label{sec:linresp}
\subsection{Derivation of $\Pi^\mu_\nu$}

In this Section we derive the polarization tensor, including damping, for the idealized case where the QGP is homogeneous and stationary. We follow the derivation presented in \cite{Formanek:2021blc} for the damped polarization tensor of an electron-positron plasma. The calculation differs slightly from \cite{Formanek:2021blc}, since in QGP we consider three quark species: up, down, and strange. We start from the Vlasov-Boltzmann equation for each quark flavor:
\begin{equation}\label{eq:VBE}
(p \cdot \partial) f_f(x,p) + q_f F^{\mu\nu} p_\nu \frac{\partial f_f(x,p)}{\partial p^\mu} = (p\cdot u)C_f(x,p)\,,
\end{equation}
The collision term $C_f(x,p)$ in the BGK form is given by
\begin{equation}\label{eq:collision}
    C_f(x,p) =\kappa_f\left(\eq{f}_f (p)\frac{n_f(x)}{{\eq{n}_f}} - f_f(x,p)\right)\,,
\end{equation}
where plasma constituents collide on a momentum-averaged time scale $\tau_{\text{rel}} = \kappa^{-1}$. The collision term is constructed such that \req{eq:VBE} retains current conservation \cite{Bhatnagar:1954zz}. We show in Sect.~\ref{sec:energymomcons} that energy is also conserved for the case of a neutral particle-antiparticle plasma at linear order in the external field.

The induced current $ j_{\mathrm{ind}}^\mu$ can be written in terms of the phase-space distribution of quarks and anti-quarks as
\begin{multline}\label{eq:current}
   j_{\mathrm{ind}}^\mu(x) = 2 N_c \int (dp)p^\mu \\ \times \sum_{u,d,s} q_f (f_{f}(x,p) - f_{\bar{f}}(x,p))\,,
\end{multline}
where  $N_c$ is the number of colors, and we sum over the quark flavors with charges $q_f$. One can calculate the induced current for small perturbations away from equilibrium for each quark flavor
\begin{equation}\label{eq:perturbation}
f_f(x,p) = {\eq{f}_f}(p) + \delta f_f(x,p)\,,
\end{equation}
Note that the equilibrium contributions ${\eq{f}_f}(p)$ do not contribute to \req{eq:current} because of the opposite sign of the charges of particles and antiparticles, but the perturbations $\delta f$ add up due the change in sign of the external force $qF^{\mu\nu}p_\nu$:
\begin{multline}\label{eq:current2}
    j_{\mathrm{ind}}^\mu(x) = 2 N_c \int (dp)p^\mu \sum_{u,d,s} q_f (\delta f_{f}(x,p) - \delta f_{\bar{f}}(x,p))\\
 = 4 N_c \int (dp)p^\mu \sum_{u,d,s} q_f^2 \delta f(x,p)\\
  =  4 N_Q e^2 \int (dp)p^\mu \delta f(x,p)\,.
\end{multline}
In the second line we pulled out a factor of electric charge $\delta f_{f} = q_f \delta f$. The perturbations $\delta f$ are identical for all quark species in the ultrarelativistic limit. The result \req{eq:current2} differs from that found in the case of an electron-positron plasma by the factor
\begin{equation}
N_Q \equiv N_c\sum_f (q_f/e)^2 = 2\,,
\end{equation}
where the numerical value holds for three light quarks flavors ($u,d,s$). We refer to \cite{Formanek:2021blc} for the derivation of the polarization tensor in terms of integrals over the phase-space distribution $\delta f$, because the only difference is the overall factor $N_Q$.

As noted in the previous Section, the polarization tensor in \req{eq:poltensgen} may be written in terms of two independent components: the longitudinal polarization function $\Pi_{\parallel}$, which describes response parallel to wave-vector $\boldsymbol{k}$, and the transverse polarization function $\Pi_{\perp}$, which describes response in the plane perpendicular to wave-vector $\boldsymbol{k}$. When the $\mu=3$ ($z$) axis is chosen along the wave-vector $\boldsymbol{k}$, the longitudinal and transverse polarization functions relate to the components of the polarization tensor \req{eq:poltenmat} along the coordinate axes as
\begin{equation}\label{eq:piLT}
    \Pi_{\parallel} =\Pi^3_3, \quad \Pi_{\perp} =\Pi^1_1=\Pi^2_2\,.
\end{equation}
In the ultrarelativistic limit, neglecting quark masses, one finds \cite{Formanek:2021blc}:
\begin{align}\label{eq:polfuncs}
&\Pi_{\parallel}(\omega,|\boldsymbol{k}|) = m_D^2\frac{\omega^2}{\boldsymbol{k}^2}\left(1 - \frac{\omega \Lambda}{2|\boldsymbol{k}|-i\kappa \Lambda}\right)\,,\\
&\Pi_{\perp}(\omega,|\boldsymbol{k}|) = \frac{m_D^2\,\omega}{4 |\boldsymbol{k}|}\left( \Lambda \left(\frac{\omega'^2}{\boldsymbol{k}^2} - 1\right) - \frac{2\omega'}{ |\boldsymbol{k}|}\right)\,,
\end{align}
where $\Lambda(\omega,\boldsymbol{k})$ is defined as
\begin{align}\label{eq:definitions}
 \Lambda \equiv \ln \frac{\omega'+  |\boldsymbol{k}|}{\omega'- |\boldsymbol{k}|}\,, \quad \text{with} \quad \omega' = \omega+i\kappa.
\end{align}
The natural logarithm leads to branch cut in the complex $\omega$ plane running from $-|\boldsymbol{k}|-i\kappa$ to  $|\boldsymbol{k}|-i\kappa$ as noted in \cite{Romatschke:2015gic}. The parallel and transverse polarization functions have the same form as in \cite{Formanek:2021blc} except for an overall factor $N_Q$ that is contained in the leading order electromagnetic Debye mass for the QGP plasma \cite{Kapusta:1992fm}:
\begin{equation}\label{eq:Debyem}
    {m_D}^2_{(\text{EM})} = \sum_{u,d,s} q^2_f T^2 \frac{N_c}{3} = N_Q\frac{e^2T^2}{3} \equiv C_{\text{em}}T^2\,,
\end{equation}
where $C_{\text{em}} =  2e^2/3$. In the following we will use $m_D$ as short-hand notation for the electromagnetic screening mass since we do not discuss color screening here.

The polarization tensor may be written in any general frame by using \req{eq:poltensgen}, but for our purposes it will be simpler to carry out calculations in the coordinate system where $\boldsymbol{k}$ aligns with the $z$-axis so that the polarization tensor takes the form shown in \req{eq:poltenmat}.

 \subsection{QGP parameters}
 
The strength of the plasma response to an external magnetic field depends on the values of two physical parameters: the quark damping rate $\kappa$, and the electromagnetic screening mass $m_D$. In this Section we provide estimates for these parameters. 

We adopt the perturbative result \req{eq:Debyem} to estimate $m_D$. Higher-order corrections to this expression can been derived from higher-order calculation of the vector spectral function in thermal perturbation theory (see \cite{Jackson:2019mop} and references cited therein).

The scale of the collisional quark damping $\kappa$ is much larger than the electromagnetic Debye mass $m_D$ because it depends on the strong coupling constant $\alpha_s$, not the electromagnetic coupling $\alpha$. Solving the dispersion relation
\begin{equation}
    \frac{1}{(k\cdot u)^2}(k^2+ \mu_0\Pi_\parallel(\omega, k))(k^2 + \mu_0 \Pi_\perp(\omega, k))^2=0 \,,
\end{equation}
see \cite{Melrose:2008}, in the limit $\boldsymbol{k}\rightarrow 0$ one finds for the plasma oscillation frequency \cite{Formanek:2021blc}
\begin{equation}\label{eq:plasmafreq}
    \omega_{p}^\pm = -\frac{i\kappa}{2} \pm \sqrt{\frac{m_D^2}{3} - \frac{\kappa}{4}^2}\,.
\end{equation}
We see that if $\kappa > \tfrac{2}{\sqrt{3}}m_D $, the plasma oscillations are over-damped.

\begin{figure}[h!]
    \centering
    \includegraphics[width=0.95\linewidth]{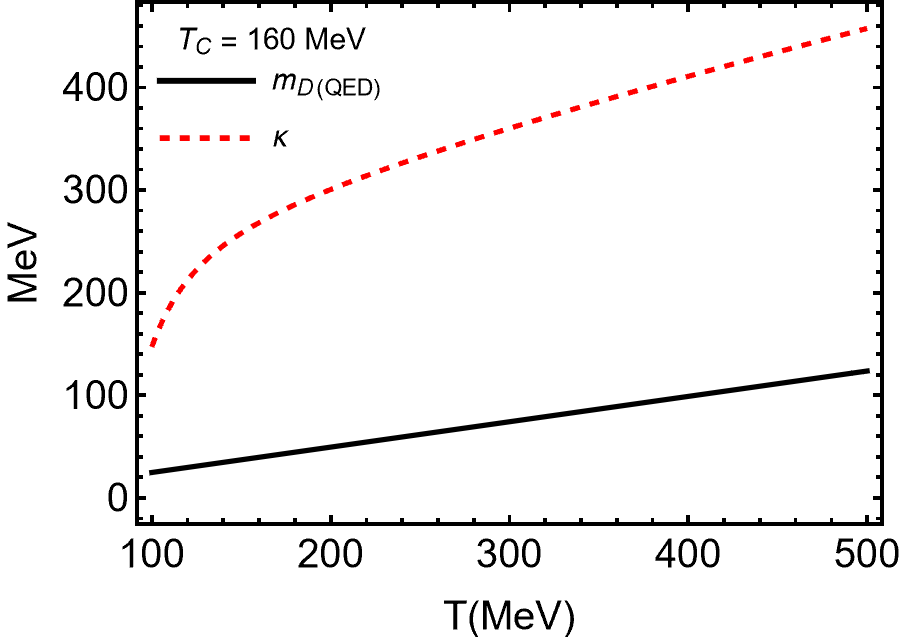}
    \caption{Plot of the QED Debye mass and the QCD dampening rate $\kappa$ as a function of temperature. At temperature $T=300\,$MeV used in the plots below, $\kappa = 4.86\, m_D$.\label{fig:kappaDebye}}
\end{figure}

The collision rate $\kappa$ is related to the inverse of the mean-free time of quarks in QGP. In kinetic theory the mean-free time is given by the product of the parton density in the QGP and the quark-parton transport cross section, leading to the expression \cite{Mrowczynski:1988xu}
\begin{equation}\label{eq:kappadef}
    \kappa(T) = \frac{10}{17\pi} (9 N_f +16) \zeta(3) \alpha_s^2 \ln\left(\frac{1}{\alpha_s}\right) T\,,
\end{equation}
where $N_f$ is the number of flavors, $\zeta(x)$ denotes the Riemann zeta function, and $\alpha_s(T)$ is the running QCD coupling.  We model the running of the QCD coupling constant as a function of temperature in the range $T<5T_c$ using a fit provided in \cite{Rafelski:2002}:
\begin{equation}\label{eq:alphas}
    \alpha_s(T) \approx \frac{\alpha_s(T_c)}{1+C \ln(T/T_c)}\,,
\end{equation}
where $C=0.760 \pm 0.002$. For the QCD (pseudo-)critical temperature we use $T_c = 160\,$MeV. The QED Debye mass is compared to $\kappa(T)$ in Fig.~\ref{fig:kappaDebye}. 

From $\kappa(T)$ in \req{eq:kappadef} and the running of the coupling in \req{eq:alphas}, we calculate the static conductivity using the leading order electromagnetic Debye mass $m_D$. The momentum dependent transverse conductivity $\sigma_{\perp}$, which controls the response of the plasma to magnetic fields, is related to the imaginary part of the transverse polarization function $\Pi_{\perp}$ as follows \cite{Melrose:2008}:
\begin{equation}\label{eq:conddef}
    \sigma_{\perp}(\omega,\boldsymbol{k}) = -i \frac{\Pi_{\perp}(\omega,\boldsymbol{k})}{\omega}\,.
\end{equation}
In the long wavelength limit $\boldsymbol{k}\rightarrow0$, the branch cut in \req{eq:definitions} shrinks to a single pole at $\omega = -i \kappa$, and the conductivity has the simple form
\begin{equation}\label{eq:conddrude}
    \sigma_{\perp}(\omega, 0 ) = \sigma_{\parallel}(\omega, 0 ) = \frac{\sigma_0}{1-i\omega/\kappa}\,.
\end{equation}
We will refer to $\sigma_{\perp}(\omega, 0 )$ as the Drude model \cite{Drude:1900}. In the static limit $\omega\rightarrow0$ the parallel and perpendicular conductivities are the same, and the static conductivity $\sigma_0$ is given by
\begin{equation}\label{eq:condstat}
   \sigma_0 = \frac{m_D^2}{3\kappa}\,.
\end{equation} 
 The static conductivity determines the late time behavior of the magnetic field.

%%%%%%%%%%%%%%%%%%%%%%%%%%%%%%%%%%%%%%%%%%%%%%%%%%%%%%%%%%%%%%%%%%%%%%%%%%%%%%%%%%%%%%%%%

\subsection{Energy-momentum conservation}\label{sec:energymomcons}

In general, the modified BGK collision term \req{eq:collision} violates energy and momentum conservation. Rocha {\it et al.} \cite{Rocha:2021zcw} recently showed how energy-momentum conservation can be restored by introducing a linearized collision operator that is projected on eigenfunctions of the conserved quantities with eigenvalue zero. Here we show explicitly that for a symmetric particle-antiparticle plasma the energy momentum violations cancel at linear order in the external field. 

Recall that the energy momentum tensor $T^{\mu\nu}$ of the plasma is given by
\begin{equation}
    T^{\mu \nu} = 2 \int (dp) p^{\mu} p^{\nu} (f_-(x,p) + f_+(x,p) )\,,
\end{equation}
where the factor of two accounts for spin and $f_\pm(x,p)$ represent the distributions of particles ($+$) and antiparticles ($-$), respectively. We recall that for $T^{\mu\nu}$ to be conserved the covariant divergence 
\begin{equation}
    \partial_{\mu} T^{\mu \nu} = 2\,\partial_{\mu} \int (dp)p^{\mu} p^{\nu}\left(f_-(x,p) + f_+(x,p)\right)
\end{equation}
must vanish. In linear response the distribution functions $ f_{\pm} (x,p)$ are given by
\begin{equation}
    f_{\pm} (x,p) = \delta  f_{\pm} (x,p) + \eq{f}(p)\,.
\end{equation}
Equation \req{eq:current2} indicates that the perturbation $ \delta  f_{\pm}$ is linear in the quark charge
\begin{equation}
    \delta f_\pm = \pm q\, \delta f\,.
\end{equation}
This leads to a cancellation of the particle and antiparticle perturbations in the energy-momentum tensor at linear order:
\begin{equation}
    \partial_{\mu} T^{\mu \nu} = 4\,\partial_{\mu}\left( \int (dp)p^{\mu} p^{\nu}\eq{f}(p)\right) = 0\,.
\end{equation}
Thus for a symmetric particle-antiparticle plasma corrections to the energy-momentum tensor appear only at second order in external field. This is a general consequence of CPT symmetry of the medium.

%%%%%%%%%%%%%%%%%%%%%%%%%%%%%%%%%%%%%%%%%%%%%%%%%%%%%%%%%%%%%%%%%%%%%%%%%%%%%%%%%%%%%%%%%%%%%%%%%%%

\section{Magnetic field in a nuclear collision}\label{sec:results}

In this Section we calculate the magnetic field at the center of the heavy ion collision by Fourier transforming the momentum space magnetic field \req{eq:ftfields} to position space. We calculate the self-consistent magnetic field using the potentials given in \reqs{eq:phi}{eq:aperp} and model the response of QGP using the idealized case of a homogeneous, stationary plasma detailed in Sects.~\ref{sec:Maxwell2} and \ref{sec:linresp}. The external fields are specified by the moving Gaussian charge distributions defined in \reqs{eq:rhoext}{eq:jext}.

The magnetic field is of particular interest due to its role in the separation of electric charge in the QGP through the chiral magnetic effect (CME) \cite{Kharzeev:2007jp}. In the large magnetic fields that occur in heavy ion collisions the electric current generated by the CME could lead to a charge separation along the direction of the magnetic field. Whether this effect is observable depends strongly on the size of the magnetic field. If a magnetic field of meaningful strength survives until the time of hadronization of the QGP, it could also lead to a difference in the global polarization of $\Lambda$ hyperons and antihyperons \cite{Muller:2018ibh}.

We chose the collision center as the origin of our spatial coordinate system and align the spatial $z$-axis with the beam direction. We calculate the magnetic field along the $z$-axis by Fourier transforming the momentum space expressions given in \reqs{eq:aperp}{eq:ftfields}:
\begin{multline}\label{eq:magorgin}
   \boldsymbol{B}(t, z) = \int \frac{d^4k}{(2\pi)^4}  e^{-i\omega t+ik_z z}
 \frac{\mu_0 i \boldsymbol{k} \times\ft{j}_{\perp \text{ext}}(\omega, \boldsymbol{k})}{\boldsymbol{k}^2 - \omega^2 - \mu_0 \Pi_{\perp}(\omega, \boldsymbol{k})}
\end{multline}
to position space. It is convenient to perform the Fourier integrals in cylindrical coordinates $(\boldsymbol{x}_\perp,z)$. The angular integral $d\theta$ and the integral over momentum along the beam axis $d k_z $ can be performed exactly. The $d k_z $ integral is trivial due to the delta function in the external charge distribution \req{eq:extchgfreq}. The frequency integral $d\omega$ and the transverse momentum integral $dk_\rho$ must, in general, be done numerically. We present the details of this calculation in Appendix \ref{sec:magf}. Due to the symmetry of the colliding ions, the only nonzero component of the magnetic field along the $z$-axis points out of the collision plane ($x-y$ plane). In our coordinate system, described in Appendix \ref{sec:freechg}, this corresponds to the $y$-component of the magnetic field. The numerical results for the position-space magnetic field are shown in Fig.~\ref{fig:bfcomp} and compared with earlier results.

To connect to these previous studies, we compute the magnetic field in position space at the origin in various levels of approximation defined in \reqs{eq:conddef}{eq:condstat} and \req{eq:lightcone}.
\begingroup
\renewcommand{\arraystretch}{1.5} % Default value: 1
\begin{table}[b]
\caption{\label{tab:cond}
Conductivity models used to calculate the resulting magnetic field. Each conductivity represents the response of QGP with a different spacetime dependence. }
\begin{ruledtabular}
\begin{tabular}{ccc}
\textrm{Conductivity}&
\textrm{Dependence}&
\multicolumn{1}{c}{\textrm{Formula}}\\
\colrule
 \textrm{Full} & $\sigma_{\perp}(\omega,\boldsymbol{k})$ & $-i \Pi_{\perp}(\omega,\boldsymbol{k})/\omega$  \\
 \textrm{Light-cone} & $\sigma_{\perp}(\omega = |\boldsymbol{k}|)$ & \req{eq:lightcone}  \\
\textrm{Drude} & $\sigma_{\perp}(\omega, 0 )$ & $\sigma_0/(1-i\omega/\kappa)$ \\
\textrm{Static} & $\sigma_0$ & $m_D^2/(3\kappa)$ \\
\end{tabular}
\end{ruledtabular}
\end{table}
\endgroup
These conductivities, collected in Table\,\ref{tab:cond}, refer to different treatments of the frequency $\omega$ and wave-vector $\boldsymbol{k}$ dependence of the conductivity $\sigma_\perp(\omega, \boldsymbol{k})$. For instance, solving for the magnetic field in the limit $\boldsymbol{k}\rightarrow0$ assumes that the spatial dependence of the external field can be neglected, not superficially a good approximation because at any given time $t$ the field varies rapidly with $z$.  The levels of approximation we consider include: the full space- and time-dependence of the conductivity $\sigma_\perp(\omega, \boldsymbol{k})$, the Drude model \req{eq:conddrude}, and the static response $\sigma_\perp(0,0)$. We list these limits in \reqs{eq:conddef}{eq:condstat}, respectively. 

The fourth limit we are considering is the conductivity along the light-cone $\sigma_\perp(\omega,k_z=\pm\omega,k_\rho=0)$. We now show that the light-cone limit closely resembles the Drude model. We first recall that the frequency dependence of the free charge distribution in cylindrical coordinates \req{eq:extchgfreq} has the form
\begin{multline}
\wt{\rho}_{\text{ext}\pm}(\omega,\boldsymbol{k}) = 2\pi Zq\, e^{-(k_{\rho}^2 + k_z^2/\gamma^2)\frac{R^2}{4}} \\
\times e^{\mp \frac{ik_{\rho} b \cos\theta }{2}} \delta(\omega \mp k_z \beta)\,.
\end{multline} 
After performing the Fourier transform over the parallel component of the wave-vector $k_z$ using the delta function, the magnitude of the wave-vector $|\boldsymbol{k}|$ is effectively set to the light-cone $\omega \approx |\boldsymbol{k}|$, with a small deviation due to the transverse dependence of the field,
\begin{equation}
    |\boldsymbol{k}|^2 = k_z^2+ k_{\rho}^2 \rightarrow  (\omega/\beta)^2+ k_{\rho}^2\,.
\end{equation}
Inspecting the external charge distribution after this replacement
\begin{multline}
\wt{\rho}_{\text{ext}\pm}(\omega,k_{\rho}) = 2\pi Zq\, e^{-(k_{\rho}^2 + \omega^2/(\beta \gamma)^2)\frac{R^2}{4}} \\
\times e^{\mp \frac{ik_{\rho} b \cos\theta }{2}}\,,
\end{multline}
we can see that the size of the deviation from the light-cone due to $k_{\rho}$ is or order $O(1/R)$, while the width of the current distribution in frequency space is of order $O(\beta\gamma/R)$.
\begin{figure}[h!]
    \centering
    \includegraphics[width=0.95\linewidth]{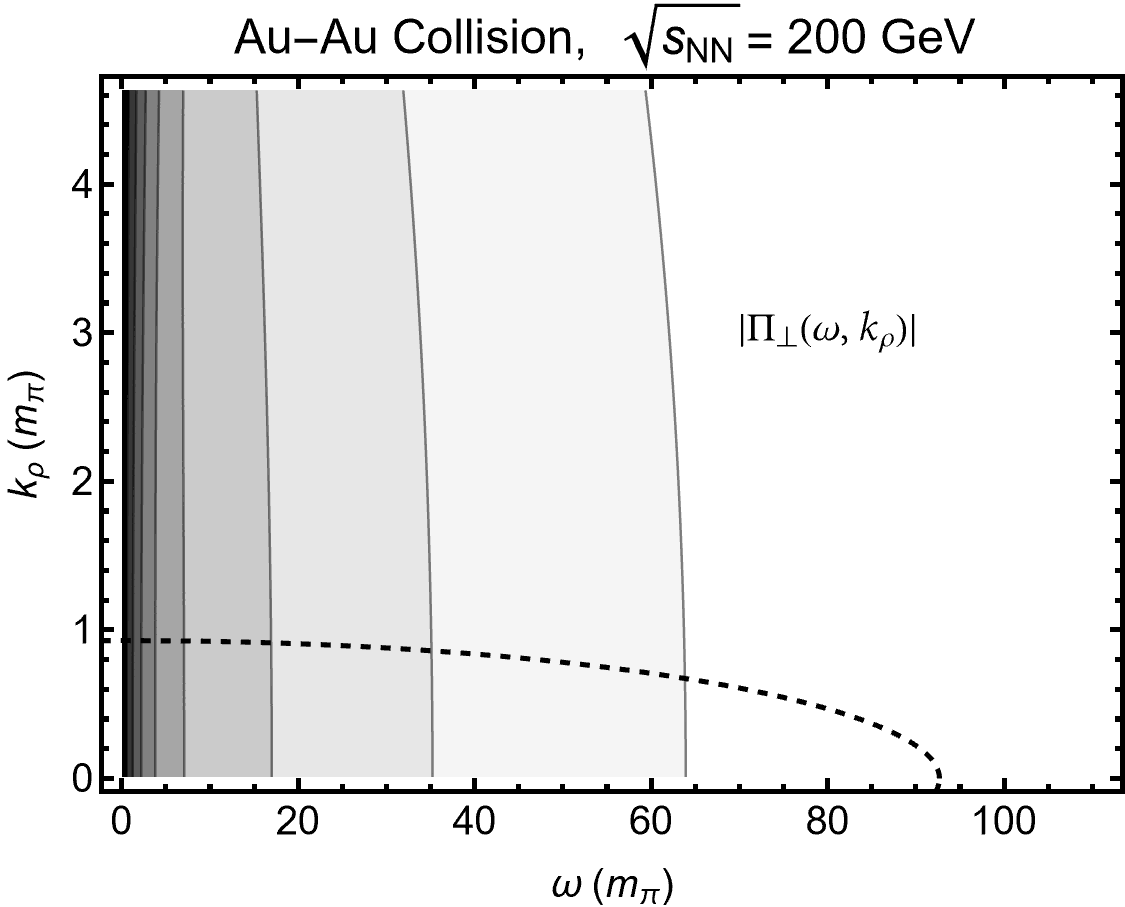}
    \caption{The magnitude of the polarization tensor is plotted in momentum space showing deviations in $k_\rho$ from the light-cone $\omega = |\boldsymbol{k}|$ on the horizontal axis. The contours show lines of constant magnitude of $\Pi_\perp(\omega, |\boldsymbol{k}|)$; lighter shading indicates increasing magnitude. The dashed line encapsulates the $2\sigma$ support of the external charge distribution. The width of the external charge distribution in momentum space is $\sqrt{2}/R$ in the transverse direction and  $\beta \gamma \sqrt{2}/R$ along the light-cone. One can see that in the region sampled by the external charge distribution the polarization tensor is effectively constant as a function of $k_\rho$. \label{fig:lightcone}}
\end{figure} 
The region of two-sigma support of the Gaussian charge distribution is shown as the region enclosed by the dashed line in Fig.~\ref{fig:lightcone}. The polarization tensor is approximately constant as a function of $k_\rho$ in this region. This implies that one can approximate the integral in \req{eq:magorgin} using the polarization function at $k_\rho = 0$, i.~e.\, on the light-cone. This means that the fields of the ions, traveling near the speed of light, probe the polarization tensor along the light-cone. In this limit, the transverse conductivity near the light-cone is
\begin{equation}\label{eq:lightcone}
    \sigma_\perp (\omega = |\boldsymbol{k}|)  =  i \frac{m_D^2}{4 \omega}\left( \frac{\kappa^2}{\omega^2} \xi \ln\xi +\frac{i\kappa}{\omega}\left(\xi+1\right)\right)\,,
\end{equation}
where $\xi$ is defined as
\begin{equation}\label{eq:xidef}
    \xi \equiv 1- 2i \frac{\omega}{\kappa}\,.
\end{equation}
Since the light-cone conductivity only depends on a single variable ($\omega = |\boldsymbol{k}|$) it simplifies integrals involved in the Fourier transform of fields back into position space.

Our results for the magnetic field at the collision center $B_y(t,0)$ are shown in Fig.~\ref{fig:bfcomp}. The right panel of the figure shows the field at early times ($|t| < 0.25~\text{fm}/c$) on a linear scale, the left panel shows the field over a wider time range on a logarithmic scale. The most general case $\sigma_\perp(\omega, \boldsymbol{k})$, shown as the dashed red curve in Fig.~\ref{fig:bfcomp}, includes the full time- and space-dependent response of the medium to the fields of the colliding ions. The blue dashed curve shows the magnetic field in the  Drude model approximation \req{eq:conddrude}, where the response depends only on time. The magnetic field using the light-cone conductivity is seen as the gray line overlapping the red dashed line in Fig.~\ref{fig:bfcomp},  where $\sigma_0$ is defined in \req{eq:condstat}. The result of Fourier transforming this expression is shown as the brown dotted curve in Fig.~\ref{fig:bfcomp}. Our results differ slightly from those of \cite{Tuchin:2013apa} because here we account for the finite size of the ions and use a slightly different conductivity value. 
\begin{widetext}
\phantom{Phantom text}
\begin{figure}[htb]
\centering
\includegraphics[width=0.46\linewidth]{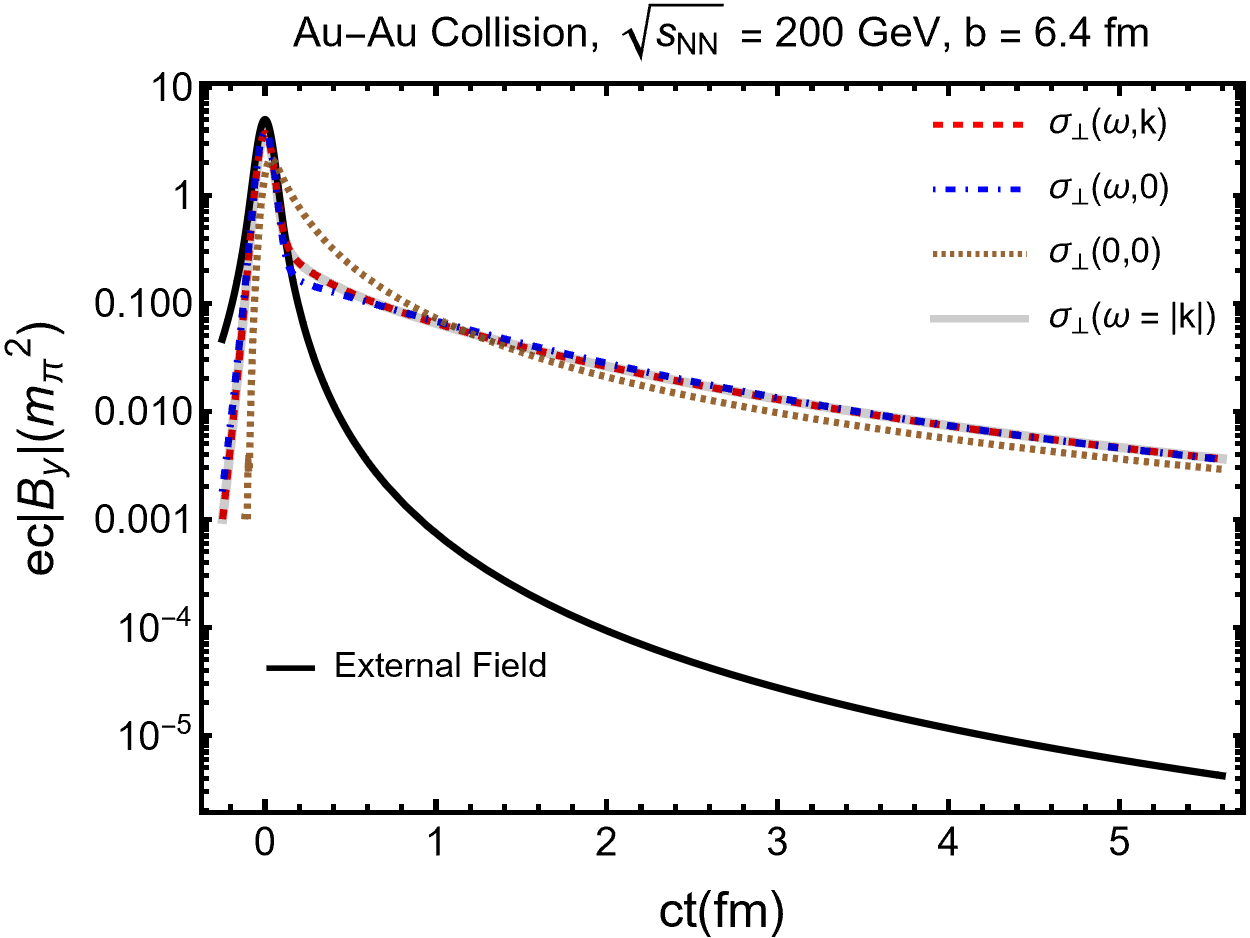}
%}
\hspace{0.05\linewidth}
\includegraphics[width=0.44\linewidth]{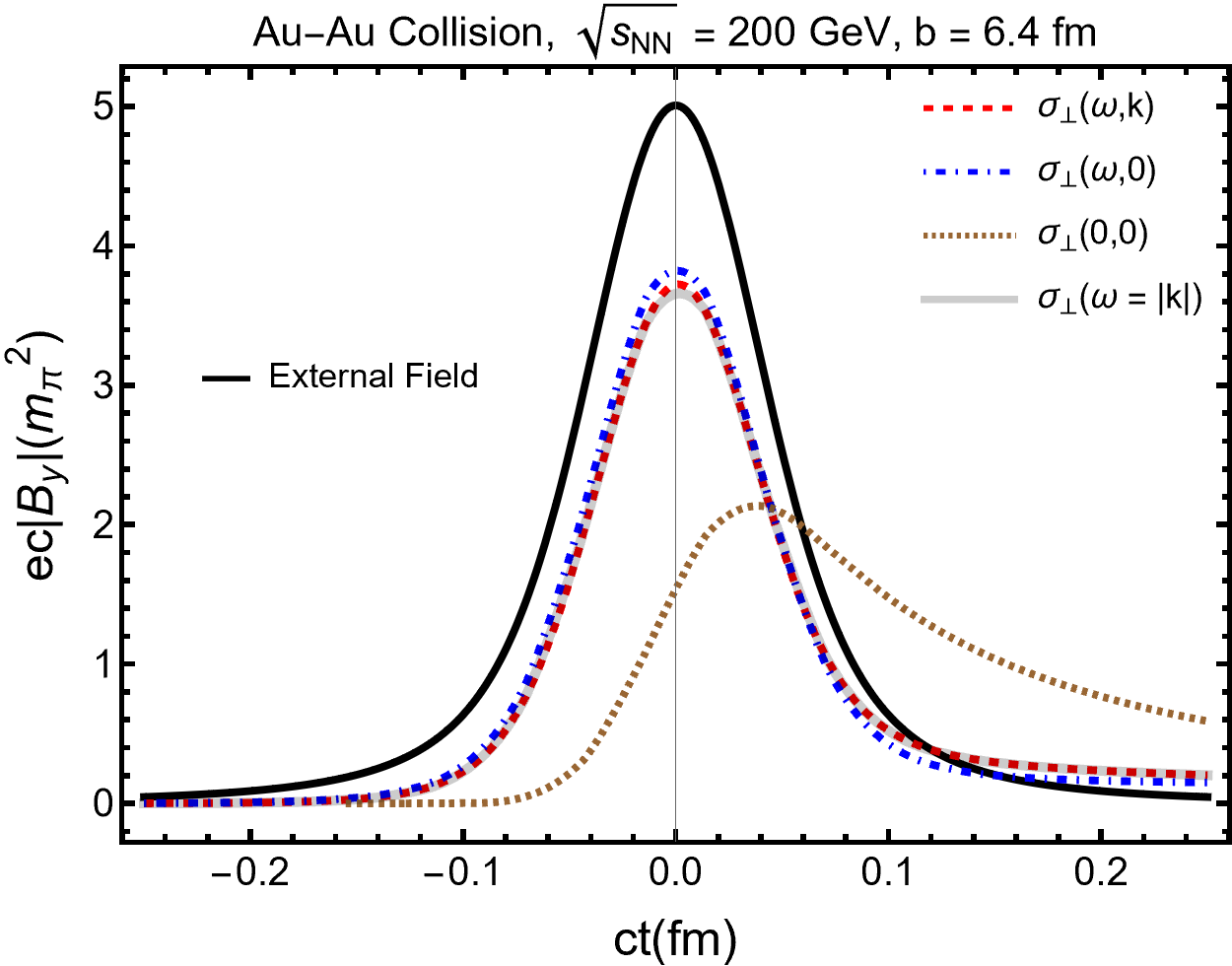}
%}
\caption{The magnetic field at the collision center as a function of time, with $T = 300$\,MeV for a Au-Au collisions ($Z=79$) at $\sqrt{s_\text{NN}} = 200$\,GeV and impact parameter $b = 6.4\,$fm. The left panel shows the magnetic field on a semi-logarithmic scale up to $ct = 5$\,fm. The right panel shows the early-time magnetic field on a linear scale. At the chosen temperature the electromagnetic Debye mass is $m_D = 74\,$MeV and the quark damping rate is $\kappa = 4.86\,m_D$. This gives a static conductivity of $\sigma_0 = 5.01\,$MeV. Comparing the different approximations we see that all of them have similar asymptotic behavior. Only the Drude conductivity, the light-cone limit of the conductivity, and the full conductivity $\sigma_\perp(\omega,\boldsymbol{k})$ describe the field at early times. Note that here that the plasma is considered homogeneous and stationary. In a more realistic situation the field would become screened only after the QGP is formed in the collision.\label{fig:bfcomp}}
\end{figure}
\end{widetext}
The magnetic field in the presence of a QGP was previously calculated using a static conductivity in \cite{Tuchin:2013apa}. In this case, the magnetic field in Fourier space has the form
\begin{equation}\label{eq:bstat}
    \ft{B}(\omega,\boldsymbol{k}) = \frac{ \mu_0 i\boldsymbol{k} \times \ft{j}_{\perp \text{ext}}}{\boldsymbol{k}^2 - \omega^2 - i\omega\sigma_0}\,,
\end{equation}

Looking at the left panel of Fig.~\ref{fig:bfcomp}, one can see that every model of the response function predicts similar values for the magnetic field approaching the freeze-out time $t_f$. This is because the static conductivity determines the late-time dependence of the magnetic field. As we discuss in Appendix \ref{sec:magf}, we can expect the static solution to match the full solution when $t > 1/\kappa$. The static conductivity initially overestimates the magnetic field after the external field begins to fall, since the effect of dynamic screening is not captured. This matches the qualitative picture given by the detailed numerical transport calculation done in \cite{Wang:2021oqq}. The full space-time dependent model and the Drude model model behave similarly for most times, and are almost identical for $t>1/\kappa \approx 0.36$\,fm/c. The approximation of the polarization tensor on the light-cone agrees very closely tracks the full solution at all times.

\begin{figure}[h!]
    \centering
    \includegraphics[width=0.85\linewidth]{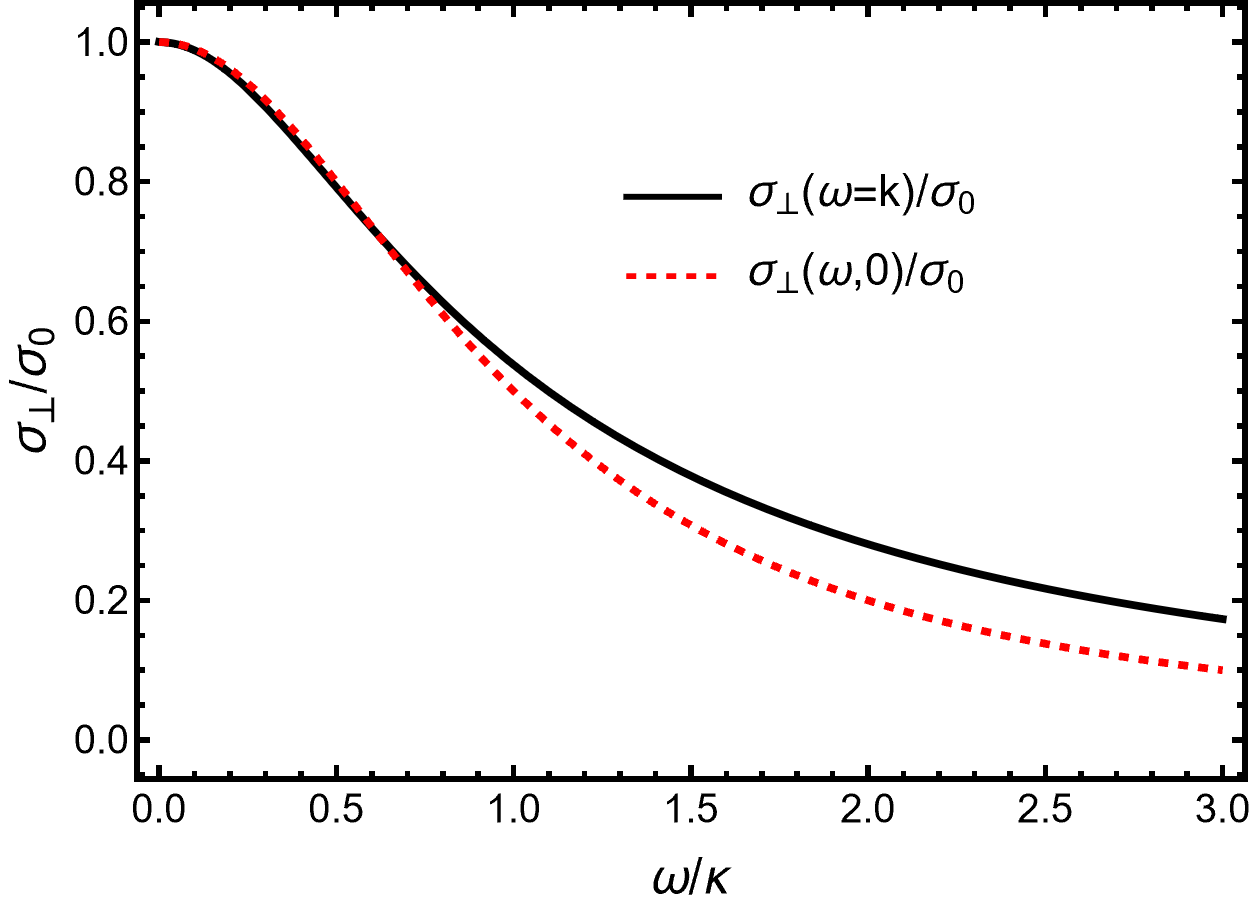}
    \caption{Comparison of the conductivity on the light-cone to $ \sigma_{\perp}(\omega,\boldsymbol{k}\rightarrow 0) $, scaled with the static conductivity. We see that at small $\omega/\kappa$, i.e. times much larger than $1/\kappa$, both approximations converge to the static case, while they diverge $\omega/\kappa > 1$. This predicts that the Drude model will underestimate screening at small times, which is exactly what we observe in Fig.~\protect\ref{fig:bfcomp}. \label{fig:condlightcomp}}
\end{figure}

We can use the light-cone conductivity in \req{eq:lightcone} to understand why the Drude model $\sigma_\perp(\omega, 0)$ matches the full solution for times $t>1/\kappa$. Late times probe the small frequency limit of the conductivity. An expansion of \req{eq:lightcone} in $\omega/\kappa$ yields
\begin{multline}
\sigma_\perp (\omega = |\boldsymbol{k}|)   = \sigma_0\left(1+i\omega/\kappa\right)\\
- \frac{6 \sigma_0}{5 }\frac{\omega^2}{\kappa^2}+O\left(\frac{\omega^3}{\kappa^3}\right)\,.
\end{multline}
We then compare to the same expansion for the Drude conductivity
\begin{multline}
\sigma_\perp (\omega,0) = \frac{\sigma_0}{1- i\omega/\kappa}  \approx  \sigma_0\left(1+i\omega/\kappa\right) \\
- \sigma_0 \frac{\omega^2}{\kappa^2}+O\left(\frac{\omega^3}{\kappa^3}\right)\,.
\end{multline}
The lowest-order term, which coincides with the expression for the Drude model, closely approximates the full solution when $\kappa \gg \omega$ as shown in Fig~\ref{fig:condlightcomp}. Since $\kappa t_f \gg 1$ for the QGP, the series converges rapidly for times of the order of the freeze-out time $t_f$. 

The simple form of the Drude approximation \req{eq:conddrude} allows one to find the poles of the denominator in \req{eq:magorgin}, analytically. The frequency integral can then be done using the residue theorem, allowing for an approximate analytical expression for the late-time magnetic field. This is done in Appendix \ref{sec:magf}. 
In the ultrarelativistic limit $\gamma\gg 1$ and large times $t\gg 1/\kappa$ gives
\begin{equation}\label{eq:banalyticapp}
   B_y(t) = -\mu_0 \frac{ Zq \beta }{(2\pi)} \frac{ b\sigma_0}{4t^2} e^{\frac{-b^2 \sigma_0}{16 t}}\,,
\end{equation}
This result differs from the ``diffusive'' solution of Tuchin \cite{Tuchin:2013apa} by a factor 1/4 in the exponent, due to their convention for impact parameter $b\rightarrow2b$. The reason why $\kappa$ does not appear in the expression \req{eq:banalyticapp} for the late-time magnetic field lies in the hierarchy of time scales $t_\text{coll} \ll 1/\kappa \ll t_f$, which makes plasma damping irrelevant during the spike of the external field as well as at freeze-out.

Interestingly, this solution has a finite limit when $\gamma\rightarrow\infty$ as it depends only on $\beta$, but not on $\gamma$. This property, which was first observed by Tuchin \cite{Tuchin:2013apa}, can be understood as follows: For late times the Fourier integral of \req{eq:extchgfreq} is dominated by contributions from small frequencies $\omega$, and it is sufficient to consider the $\omega \rightarrow 0$ limit of the Fourier spectrum of the external charge distributions $\wt{\rho}_{f\pm}$ given in \req{eq:By}. In this limit \req{eq:extchgfreq} takes the form
\begin{equation}
\wt{\rho}_{\text{ext}\pm}(0,\boldsymbol{k}) \rightarrow 2\pi Ze\, e^{-k_\rho^2R^2/4} e^{\mp \frac{i k_\rho b \cos \theta }{2}} \delta(k_z \beta)\,,
\end{equation} 
which is independent of $\gamma$. This occurs because
\begin{equation}
\wt{\rho}_{\text{ext}\pm}(0,\boldsymbol{k}) = \int dt \int d^3x e^{-\boldsymbol{k}\cdot\boldsymbol{x}} \rho_{\text{ext}\pm}(0,\boldsymbol{x})
\end{equation}
integrates over the passage of the entire nucleus at a given location $\boldsymbol{x}$ and thus is independent of $\gamma$ as the total charge is Lorentz invariant. We conclude that, quite generally, for high collision energies the remnant magnetic field at late times is determined by the time-integrated action of the external electromagnetic pulse on the QGP. In a more realistic calculation, where the QGP is not present for the entire duration of the pulse, because it is created during the collision, the remnant magnetic field will be diminished as only a fraction of the pulse acts on the QGP. We therefore expect our result to represent an upper bound to the late-time magnetic field in a realistic collision scenario.

\begin{figure}
    \centering
\includegraphics[width=0.95\linewidth]{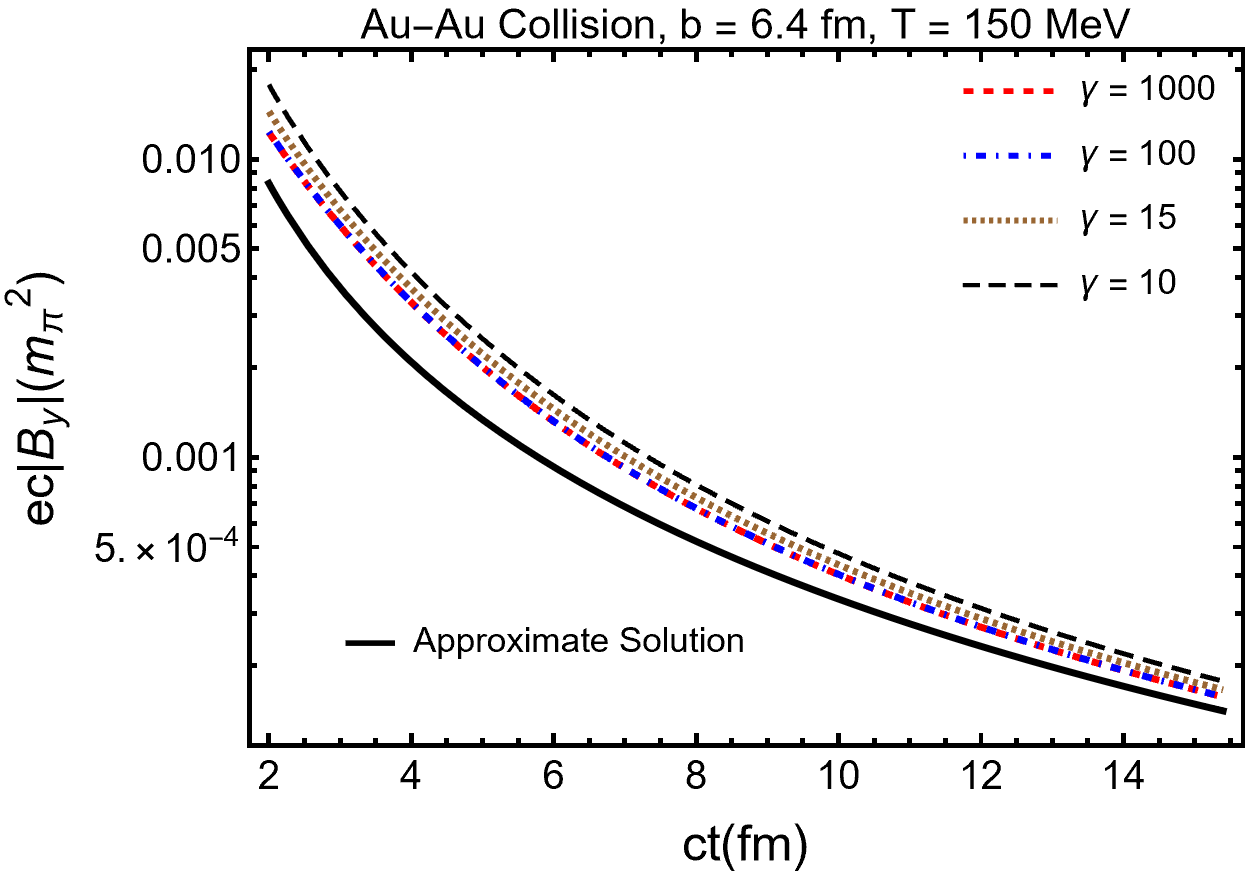}
    \caption{Plot of the freeze-out magnetic field for $T= 150$\,MeV. We expect that around this temperature QGP will hadronize into a mixed phase \cite{Letessier:1992xd}. The approximate late time solution \req{eq:banalyticapp} shown as a black line is compared to numerical calculations using the full polarization tensor \req{eq:magorgin}. the approximate solution does not fully match the ultrarelativistic limit until times $t > t_{\sigma} \approx 59$\,fm/c. The magnetic field is independent of the beam energy over a wide range of $\gamma$ but begins to diverge slowly from the ultrarelativistic case at around $\gamma \leq 15$ for the time window shown in the figure. Lower beam energies result in a somewhat larger field at late time.\label{fig:bcolcomp}}
\end{figure}

The approximation used to derive this solution holds for $\gamma\beta \gg \sqrt{ \kappa/\sigma_0} \approx 12$. In Fig.~\ref{fig:bcolcomp} we compare \req{eq:banastat} to the full numerical result to explore its dependence on $\gamma$.  One can see that the ultrarelativistic case (black solid line) begins to diverge from the numerical solution at around $\gamma \approx 15$ for the times shown. The early time magnetic field is not shown because the initial temperature of QGP will depend strongly on the collision energy. The times are chosen such that they cover the range of freeze-out times predicted for QGP for the range of experimental collision energies used \cite{Bass:2000ib}. We do not show curves for $\gamma<10$ because we expect the effects of chemical potential will become important, yet here chemical potential $\mu$ is set to zero. In Fig.~\ref{fig:bcolcomp} one can see that the late-time magnetic field has a very soft dependence on collision energy. The time at which the magnetic field freezes out, which varies with collision energy, has a much stronger effect on the magnitude of the freeze-out field.

As the QGP begins to hadronize at time $t_f$, one may expect hadrons to be statistically polarized with respect to the magnetic field. In \cite{Muller:2018ibh} the measured difference in global polarization of hyperons and antihyperons is used to give an upper bound on the magnetic field at QGP freeze-out, $B \sim 2.7\times 10^{-3}\,m_{\pi}^2$ for Au+Au collisions at $\sqrt{s_\text{NN}} = 200$\,GeV. Looking at Fig.~\ref{fig:bcolcomp} the magnetic field for $\gamma = 100$ at QGP freeze-out $t_f \approx 5 $\,fm/c is predicted to be $B \approx 1.2\times 10^{-3}\,m_{\pi}^2$, somewhat below this upper bound. Note that this assumes the polarization  rapidly equilibriates in the plasma. It also neglects any interactions during the hadron gas phase of the collision. 

\begin{figure}
\vskip 16pt
    \centering
\includegraphics[width=0.95\linewidth]{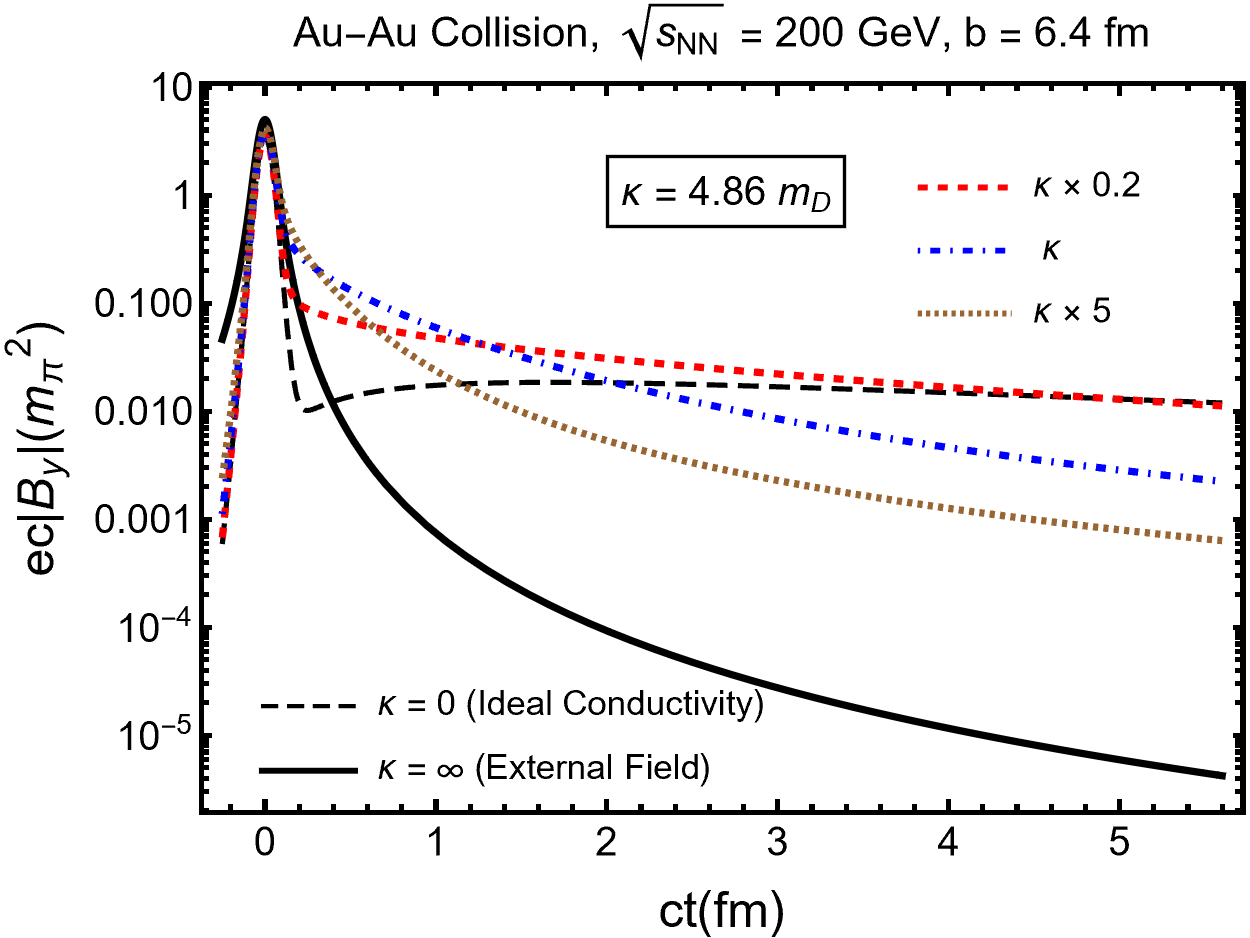}
    \caption{Comparison of the magnetic field for different values of quark damping rate or, equivalently, electric conductivity. Larger values of  the damping rate $\kappa$ represent smaller conductivities and vice versa as indicated by \req{eq:condstat}. The black dashed line and the solid black line represent the limits of zero and infinite conductivity, respectively. One can see that as $\kappa$ increases the asymptotic value of the magnetic field decreases.\label{fig:kappacomp}}
    
\end{figure}

In Fig.~\ref{fig:kappacomp} we look at the magnetic field at the origin for different values of $\kappa$. Increasing $\kappa$ reduces the static conductivity $\sigma_0$ which decreases the asymptotic value of the magnetic field as indicated by \req{eq:banalyticapp}. As $\kappa$ goes to zero the results converge to the case of ideal conductivity  $\sigma_0 \rightarrow \infty$ where the magnetic field quickly approaches a constant value. This case was studied in \cite{Deng:2012pc} where the authors considered a magnetic field that falls to a constant value and then decreases with $1/t$ due to Bjorken flow. More recent calculations \cite{Yan:2021zjc,Wang:2021oqq} solve the Vlasov-Boltzmann equation numerically with parton-parton scattering. The magnetic field predicted by \cite{Yan:2021zjc} is around $\sim 10^{-4} m_\pi^2$ after $t\approx 2$\,fm/c, which is two orders of magnitude lower than the value found here (see Fig.~\ref{fig:bcolcomp}). However, the magnetic field predicted by \cite{Wang:2021oqq} is around $\sim 10^{-2} m_\pi^2$ after $t\approx 2$\,fm/c, which is in agreement with our model.

In Fig.~\ref{fig:lighfield} we show a space-time contour plot of the magnetic field. The field at the higher collision energy (on the left) has a higher peak magnetic field. For lower collision energy (on the right) the field is less Lorentz contracted, and leads to a magnetic field at late times that is a factor of $\sim1.1$ larger. The freeze-out magnetic field will increase at lower collision energy mainly due to the decreasing freeze-out time.

\begin{widetext}
\phantom{Phantom text}

\begin{figure}[H]
\centering
\includegraphics[width=0.45\linewidth]{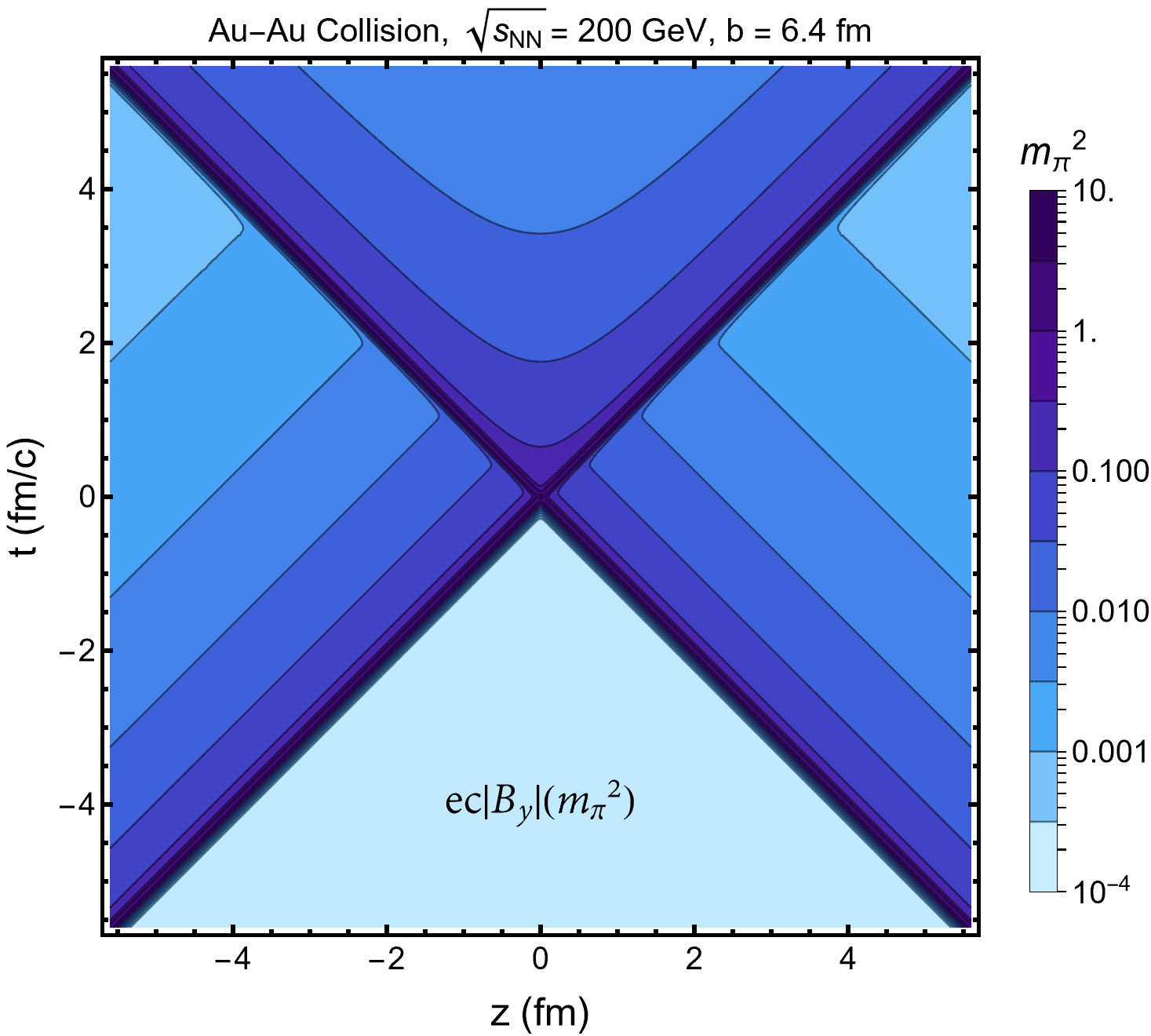}
%}
\hspace{0.05\linewidth}
\includegraphics[width=0.45\linewidth]{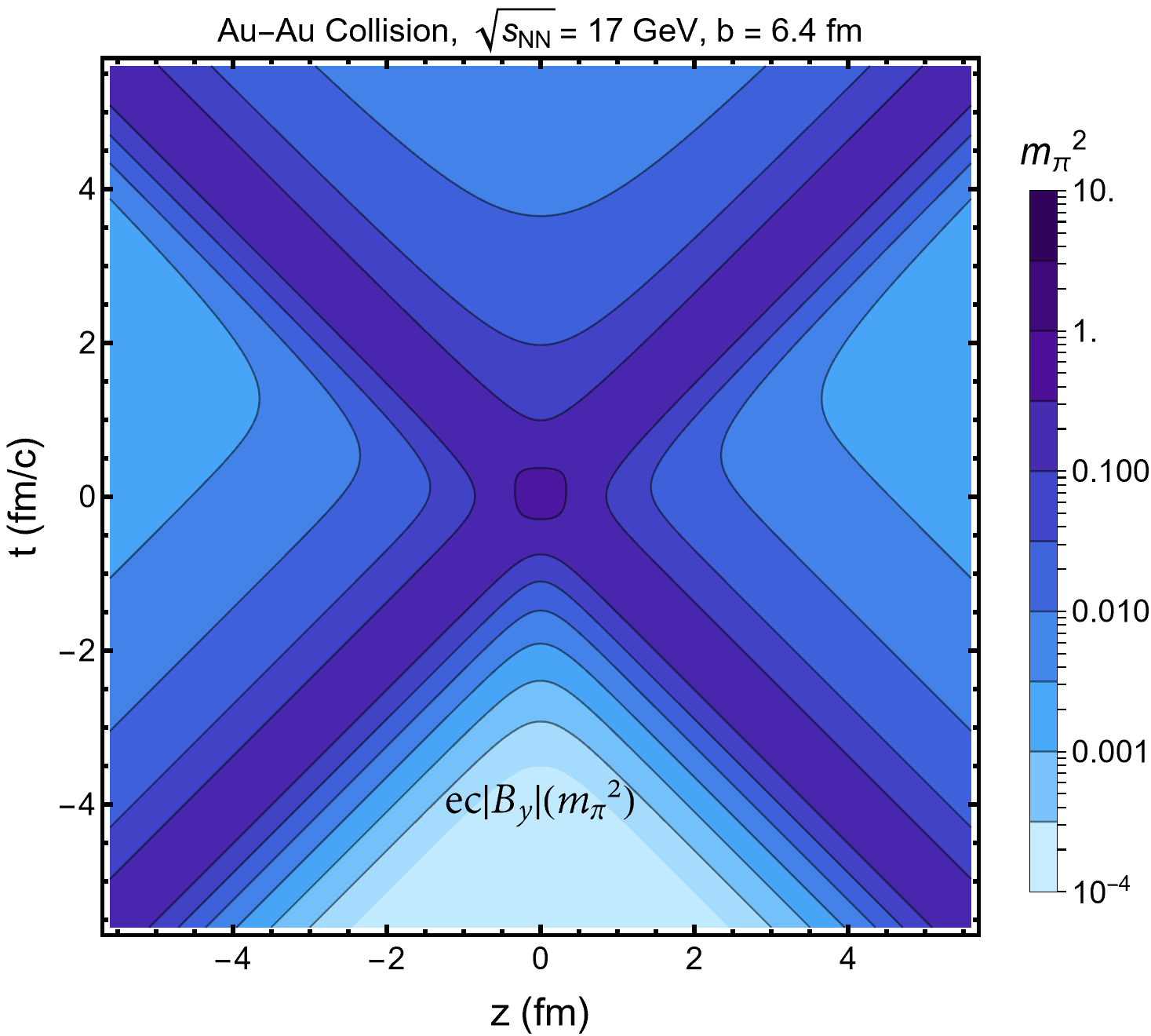}
%}
\caption{Space-time plot of the magnetic field on the beam axis ($x=y=0$) in eternal (pre-existent) QGP with $T = 300$\,MeV for a Au-Au collision at impact parameter $b = 6.4\,$fm. Left panel: collision energy $\sqrt{s_\text{NN}} = 200$\,GeV; right panel: collision energy $\sqrt{s_\text{NN}} = 17$\,GeV. The same value of $\kappa$ is used as in Fig.~\ref{fig:bfcomp}. In a more realistic scenario, where the QGP is formed during the collision, the field would only create induced currents in the upper light-cone.\label{fig:lighfield}}
\end{figure}
\end{widetext}

%%%%%%%%%%%%%%%%%%%%%%%%%%%%%%%%%%%%%%%%%%%%%%%%%%%%%%%%%%%%%%%%%%%%%%%%%%%%%%%%%%%%%%%%%%%%%%%%%%%%%%%%%%%%%%%%%%%%%%%%

\section{Summary and Conclusion}\label{sec:Conclusions}

We have studied the linear response of a stationary, homogeneous QGP to the electromagnetic field of two colliding nuclei in relativistic kinetic theory with collisional damping. As an application, we calculated the magnetic field between the two nuclei. We showed that the response to the external field is controlled by the polarization function along the light-cone, $\Pi^\mu_\nu(\omega,|\boldsymbol{k}|\approx\omega)$. This allowed us to derive an approximate analytic solution for the magnetic field that takes into account the dynamical medium response. We showed that the late-time magnetic field is mainly determined by the static electric conductivity of the QGP. Since the remnant magnetic field at hadronization does not depend strongly on the collision energy an experimental measurement of the magnetic field at different collision energies could permit a determination the electrical conductivity of the QGP. 

Our calculation can be improved in various ways to make it more realistic. Such improvements would include a realistic space-time dependence of the medium (formation and hydrodynamical evolution), nonzero net baryon density, and viscous corrections to the unperturbed phase-space distribution used to calculate the polarization tensor. These improvements will require numerical solutions of the linear response equations. It would also be of interest to study the energy-momentum deposition into the QGP by the external electromagnetic field.

Beyond heavy-ion collisions, the homogeneous polarization tensor with collisional damping studied in our work has applicability in  cosmology, where a QGP existed during the first $10~\mu$s of the early universe, and possibly in astrophysics where quark matter can exist at the core of collapsed stars. In these situations the assumption of homogeneity and stationarity of the medium on the scale of the relevant parameters, $m_D$ and $\kappa$, is well justified.

\begin{acknowledgments}
An allocation of computer time from the UA Research Computing High Performance Computing (HPC) at the University of Arizona is gratefully acknowledged. BM acknowledges support by the U.S. Department of Energy Office of Science under Grant DE-FG02-05ER41367.
\end{acknowledgments}

%===================================================================
%===================================APPENDICES======================
\appendix
%===================================================================
\section{Electric current of two colliding nuclei}\label{sec:freechg}

Here we define the free charge and current density used to describe heavy ion collisions. We wish to model two nuclei moving at constant velocity $\pm \beta$ along the collision axis ($\hat{z}$ direction) that are offset by $\pm b/2$ within the collision plane ($\hat{x}$ direction).  For simplicity we model the charge distribution as a gaussian in all directions
\begin{multline}
\rho_{\text{ext}\pm }(t,\boldsymbol{x}) = \frac{Ze\gamma}{\pi^{3/2}R^3}e^{-\frac{1}{R^2}(x\mp b/2)^2}e^{-\frac{1}{R^2}y^2}\\
\times e^{-\frac{\gamma^2}{R^2}(z\mp \beta t)^2}\,,
\end{multline}
where the normalization is chosen in such a way that 
\begin{equation}
\int \rho_{\text{ext}\pm}(t,\boldsymbol{x}) d^3\boldsymbol{x} = Ze\,,
\end{equation}
is the total charge of the heavy ion nucleus and $\gamma$ is the usual relativistic factor. The Gaussian radius parameter $R$ is related to the mean squared radius of the nucleus, $\langle r^2\rangle$ at rest, ($\gamma = 1$) by
\begin{equation}\label{eq:radius}
\langle r^2 \rangle = \frac{1}{Ze}\int r^2 \rho_{\text{ext}\pm}(\boldsymbol{x}) d^3\boldsymbol{x} = \frac{3}{2}R^2\,,
\end{equation}
which is measured experimentally for a gold nucleus to be $\sqrt{\langle r^2 \rangle} \approx 5.30 \,$fm \cite{DeVries:1987atn} .

At time $t = 0$ both nuclei are localized at the $z = 0$ plane and we assume that before and after the collision they continue moving on a straight line along the $z$-axis. The gaussian form of the charge distributions allows us to evaluate the Fourier transformations easily. The transforms in the transverse directions are
\begin{align}
\int_{-\infty}^\infty dy \ e^{-ik_y y}e^{-y^2/R^2} &= R \sqrt{\pi} e^{-k_y^2 R^2/4}\,, \\ 
\int_{-\infty}^\infty \ dx e^{-ik_x x}e^{-(x\mp b/2)^2/R^2} &=  R \sqrt{\pi}e^{-k_x^2 R^2/4}e^{\pm i k_x b /2}\,.
\end{align}
The last two integrals are a bit more complicated because they are coupled
\begin{multline}
\int_{-\infty}^\infty e^{i\omega t} \left(\int_{-\infty}^\infty e^{-ik_z z}e^{- \frac{\gamma^2}{R^2}(z^2 \pm 2z\beta t)} dz \right)\\
\times e^{-\frac{\gamma^2}{R^2}\beta^2t^2}dct = \frac{R \sqrt{\pi}}{\gamma}e^{-\frac{k_z^2R^2}{4\gamma^2}} \int_{-\infty}^\infty e^{i(\omega \pm k_z \beta)t}dt\\
= \frac{2 R\pi^{3/2}}{\gamma}e^{-\frac{k_z^2R^2}{4\gamma^2}} \delta(\omega \pm k_z \beta)\,,
\end{multline}
where delta function appears because both nuclei move at a constant velocity. Altogether the Fourier transformed charge distributions are 
\begin{multline}
\wt{\rho}_{\text{ext}\pm}(\omega,\boldsymbol{k}) = 2\pi Ze\, e^{-(k_x^2 + k_y^2 + k_z^2/\gamma^2)\frac{R^2}{4}} \\
\times e^{\mp \frac{i k_x b}{2}} \delta(\omega \mp k_z \beta)\,,
\end{multline} 
which may be written in cylindrical coordinates
\begin{multline}\label{eq:extchgfreq}
\wt{\rho}_{\text{ext}\pm}(\omega,\boldsymbol{k}) = 2\pi Ze\, e^{-(k_\rho^2 + k_z^2/\gamma^2)\frac{R^2}{4}} \\
\times e^{\mp \frac{i k_\rho b \cos \theta }{2}} \delta(\omega \mp k_z \beta)\,.
\end{multline} 
The current densities are obtained from
\begin{equation}\label{eq:extcurrent}
\ft{j}_{\text{ext}\pm}(\omega, \boldsymbol{k}) = \pm \beta \hatv{z} \wt{\rho}_{\text{ext}\pm}(\omega, \boldsymbol{k})\,.
\end{equation}
The transverse component of the current given by
\begin{multline}\label{eq:jperpext}
\ft{j}_{\perp,\text{ext}} = \ft{j}_\text{ext} - (\hatv{k} \cdot \ft{j}_\text{ext}) \hatv{k} \\= (\hatv{z} - \hat{k}_z \hatv{k})\beta(\wt{\rho}_{\text{ext}+} - \wt{\rho}_{\text{ext}-})\,.
\end{multline}

%%%%%%%%%%%%%%%%%%%%%%%%%%%%%%%%%%%%%%%%%%%%%%%%%%%%%%%%%%%%%%%%%%%

\section{Magnetic field at the collision center}\label{sec:magf}

The magnetic field inside the plasma is given in Fourier space by
\begin{equation}\label{eq:magapp}
   \ft{B} = i \boldsymbol{k} \times \wt{\boldsymbol{A}} = i \boldsymbol{k} \times \wt{\boldsymbol{A}}_{\perp}\,,
\end{equation}
where the potential $\wt{\boldsymbol{A}}$ has been projected into components transverse and longitudinal to $\boldsymbol{k}$. In the following we represent the wave-vector in cylindrical coordinates $\boldsymbol{k} = (k_{\rho} \cos \theta,k_{\rho}\sin \theta, k_{z}) $. Using the expression for the self-consistent vector potential \req{eq:aperp} we find for the magnetic field,
\begin{equation}
   \ft{B} =
 \frac{\mu_0 i \boldsymbol{k} \times\ft{j}_{\perp \text{ext}}}{\boldsymbol{k}^2 - \omega^2 - \mu_0 \Pi_{\perp}}\,.
\end{equation}
Given the definition of $\ft{j}_{\perp,\text{ext}}$ in \req{eq:jperpext} we can replace the perpendicular component of the current by its full form, adding the $\pm$ components of \req{eq:extcurrent}:
\begin{equation}
   \ft{B} =
 \frac{\mu_0 i \boldsymbol{k} \times\ft{j}_{\text{ext}}}{\boldsymbol{k}^2 - \omega^2 - \mu_0 \Pi_{\perp}}\,.
\end{equation}
We now Fourier transform this quantity back to position space in order to calculate the magnetic field at the collision center as a function of time. Due to symmetry the only nonvanishing component of the magnetic field at this location will be the $y$-component:
\begin{equation}\label{eq:By}
   \wt{B}_y = 
 \mu_0 \frac{ i k_x \beta (\wt{\rho}_{\text{ext}-}-\wt{\rho}_{\text{ext}+})}{\boldsymbol{k}^2 - \omega^2 - \mu_0 \Pi_{\perp}}\,.
\end{equation}
The Fourier transform at any point along the collision axis ($x=y=0$) is given by
\begin{equation}
  B_y(z,t) =  \int \frac{d^4k}{(2\pi)^4}  e^{-i\omega t+ ik_z z} \wt{B}_y(\omega, \boldsymbol{k})\,.
\end{equation}
In cylindrical coordinates the integral can be written as
\begin{multline}
  B_y(z,t) =  \frac{1}{(2\pi)^4}\int k_{\rho}dk_{\rho} d\omega dk_z d\theta  e^{-i\omega t+ ik_z z}\\
 \mu_0 \frac{ i \beta k_{\rho} \cos \theta (\wt{\rho}_{f-}-\wt{\rho}_{f+})}{\boldsymbol{k}^2 - \omega^2 - \mu_0 \Pi_{\perp}(\omega, |\boldsymbol{k}|)}\,.
\end{multline}
We can use the the delta function in the Fourier transformed current \req{eq:extchgfreq} to trivially perform the $k_z$ integral:
\begin{multline}
   B_y(z,t) = -\mu_0 \frac{ Ze \beta }{(2\pi)^3}  \int dk_{\rho} d\omega d\theta \  e^{-i\omega t} \\ \frac{ 2 k_{\rho}^2  \cos \theta \sin \left(  k_{\rho} \cos \theta \frac{b}{2} - \frac{\omega z}{\beta} \right)
  e^{-(k_{\rho}^2+ \omega^2/(\beta\gamma)^2)\frac{R^2}{4}}}{\omega^2/(\gamma\beta)^2 + k_{\rho}^2  - \mu_0 \Pi_{\perp}\left(\omega, \sqrt{k_{\rho}^2 + \omega^2/\beta^2} \right)}\,.
\end{multline}
We next perform the angular integration
 \begin{multline}\label{eq:intnum}
   B_y(z,t) = -\mu_0 \frac{ Ze \beta }{(2\pi)^2}  \int dk_{\rho} d\omega \  e^{-i\omega t} \\ \frac{ 2 k_{\rho}^2 J_1 \left(\frac{k_{\rho} b}{2} \right) \cos \left( \frac{\omega z}{\beta}\right)
  e^{-(k_{\rho}^2+ \omega^2/(\beta\gamma)^2)\frac{R^2}{4}}}{\omega^2/(\gamma\beta)^2 + k_{\rho}^2  - \mu_0 \Pi_{\perp}\left(\omega, \sqrt{k_{\rho}^2 + \omega^2/\beta^2} \right)}\,,
\end{multline}
where $J_1$ is a Bessel function of the first kind. The remaining integrals have to be performed numerically. From here forward we pull the factor of $\mu_0$ into the Debye mass $m_D^2$, such that the factor of $e^2$ goes to $4 \pi \alpha$ in \req{eq:Debyem}.

We can obtain an analytical expression for the Drude approximation \req{eq:conddrude} in the limit $\gamma\beta \gg \sqrt{ \kappa/\sigma_0}$, which is valid when $\gamma \gg 12$ for the values of $\sigma_0$ and $\kappa$ adopted here.  In this limit we can neglect the first term in the denominator of \req{eq:intnum}, which now takes the simple form
\begin{equation}
      k_\rho^2 - i \omega \frac{\omega_p^2}{\kappa - i\omega}\,.
\end{equation}
Note that using \req{eq:condstat} we can see $\omega_p^2 = m_D^2/3 = \kappa \sigma_0 $. The integrand of \req{eq:intnum} then has a single pole at 
\begin{equation}
\omega = -i\frac{k_\rho^2\kappa}{k_\rho^2+\omega_p^2}\,,
\end{equation}
and the frequency integral can be performed by contour integration in the lower complex plane. Consistently neglecting the term proportional to $\omega^2/(\beta\gamma)^2$ in the exponent, the integration yields:
\begin{multline}\label{eq:bintfull}
   B_y(z,t) \approx - \mu_0 \frac{ Ze \beta }{2\pi}  \int dk_{\rho}\, 2\kappa k_{\rho}^2\\
    \frac{\omega_p^2}{(k_\rho^2+\omega_p^2)^2} 
    J_1 \left(\frac{k_{\rho} b}{2} \right) e^{-k_{\rho}^2R^2/4} \\
   \cosh \left( \frac{k_\rho^2\kappa}{k_\rho^2+\omega_p^2} \frac{z}{\beta}\right)
   \exp \left( - \frac{k_\rho^2\kappa t}{k_\rho^2+\omega_p^2} \right)\,.
\end{multline}
For late times $t$ the exponential factor only samples the small $k_\rho$ region of the integrand ($k_\rho^2 < \sigma_0/t$). We can then neglect $k_\rho$ with respect to $ \omega_p$ in the integrand provided that $\sigma_0/t \ll \omega_p^2$, which is satisfied when $t \gg 1/\kappa = t_{\text{rel}} \approx 1\,\text{fm}/c$. The expression then takes the simplified form:
yielding
\begin{multline}\label{eq:bappint}
   B_y(z,t) \approx - \mu_0 \frac{ Ze \beta }{2\pi}  \int dk_{\rho}\, \frac{2k_{\rho}^2}{\sigma_0}
   J_1 \left(\frac{k_{\rho} b}{2} \right) \\
   e^{-k_{\rho}^2R^2/4} \cosh \left( \frac{k_\rho^2}{\sigma_0} \frac{z}{\beta} \right) e^{-k_\rho^2 t/\sigma_0}\,.
\end{multline}
The integral over $k_\rho$ can now be performed analytically resulting in:
\begin{equation}
   B_y(z,t) \approx - \mu_0 \frac{ Ze \beta }{2\pi}  \frac{b}{8\sigma_0}
   \left[ \frac{e^{-\frac{b^2}{16L_+}}}{L_+^2} + \frac{e^{-\frac{b^2}{16L_-}}}{L_-^2} \right]
\end{equation}
with 
\begin{equation}
L_\pm = \frac{R^2}{4} + \frac{t\pm z/\beta}{\sigma_0} \,.
\end{equation}
At the collision center $(z=0)$ and for $t \gg \sigma_0R^2/4$ our result simplifies to
\begin{equation}\label{eq:banastat}
   B_y(0,t) \approx - \mu_0 \frac{ Ze \beta }{2\pi}  \frac{b\sigma_0}{4t^2} e^{-\frac{\sigma_0b^2}{16t}}\,.
\end{equation}

This result differs from Tuchin's \cite{Tuchin:2013apa} by a factor 1/4 in the exponent due to their convention for impact parameter $b\rightarrow2b$.

%%%%%%%%%%%%%%%%%%%%%%%%%%%%%%%%%%%%%%%%%%%%%%%%
%%%%%%%%%%%%%%%%%%%%%%%%%%%%%%%%%%%%%%%%%%%%%%%%

% % %%%%%%%%%%%%%%%

\begin{thebibliography}{99}

%%%%%%%%%%%%%%%%%  Analytic - Constant Conductivity 

    \bibitem{Tuchin:2010vs}
    K.~Tuchin,
    %``Synchrotron radiation by fast fermions in heavy-ion collisions,''
    Phys. Rev. C \textbf{82}, 034904 (2010)
    [erratum: Phys. Rev. C \textbf{83} (2011), 039903]
    % doi:10.1103/PhysRevC.83.039903
    [arXiv:1006.3051 [nucl-th]].
    
    \bibitem{Deng:2012pc}
    W.~T.~Deng and X.~G.~Huang,
    %``Event-by-event generation of electromagnetic fields in heavy-ion collisions,''
    Phys. Rev. C \textbf{85}, 044907 (2012)
    % doi:10.1103/PhysRevC.85.044907
    [arXiv:1201.5108 [nucl-th]].
    
    \bibitem{Tuchin:2013apa}
    K.~Tuchin,
    %``Time and space dependence of the electromagnetic field in relativistic heavy-ion collisions,''
    Phys. Rev. C \textbf{88}, 024911 (2013) 
    % doi:10.1103/PhysRevC.88.024911
     [arXiv:1305.5806 [hep-ph]].
    
    \bibitem{McLerran:2013hla}
    L.~McLerran and V.~Skokov,
    %``Comments About the Electromagnetic Field in Heavy-Ion Collisions,''
    Nucl. Phys. A \textbf{929}, 184 (2014)
    % doi:10.1016/j.nuclphysa.2014.05.008
    [arXiv:1305.0774 [hep-ph]].
    
    \bibitem{Gursoy:2014aka}
    U.~Gursoy, D.~Kharzeev and K.~Rajagopal,
    %``Magnetohydrodynamics, charged currents and directed flow in heavy ion collisions,''
    Phys. Rev. C \textbf{89}, 054905 (2014) 
    % doi:10.1103/PhysRevC.89.054905
    [arXiv:1401.3805 [hep-ph]].

    \bibitem{Roy:2015kma}
    V.~Roy, S.~Pu, L.~Rezzolla and D.~Rischke,
    %``Analytic Bjorken flow in one-dimensional relativistic magnetohydrodynamics,''
    Phys. Lett. B \textbf{750}, 45 (2015)
    % doi:10.1016/j.physletb.2015.08.046
    [arXiv:1506.06620 [nucl-th]].
    
    \bibitem{Li:2016tel}
    H.~Li, X.~l.~Sheng and Q.~Wang,
    %``Electromagnetic fields with electric and chiral magnetic conductivities in heavy ion collisions,''
    Phys. Rev. C \textbf{94}, 044903 (2016)
    % doi:10.1103/PhysRevC.94.044903
    [arXiv:1602.02223 [nucl-th]].

%%%%%%%%%%%%%%%%%  Numerical (MDH) - Constant Conductivity 

    \bibitem{Inghirami:2016iru}
    G.~Inghirami, L.~Del Zanna, A.~Beraudo, M.~H.~Moghaddam, F.~Becattini and M.~Bleicher,
    %``Numerical magneto-hydrodynamics for relativistic nuclear collisions,''
    Eur. Phys. J. C \textbf{76}, 659 (2016)
    % doi:10.1140/epjc/s10052-016-4516-8
    [arXiv:1609.03042 [hep-ph]].
    
    \bibitem{Inghirami:2019mkc}
    G.~Inghirami, M.~Mace, Y.~Hirono, L.~Del Zanna, D.~E.~Kharzeev and M.~Bleicher,
    %``Magnetic fields in heavy ion collisions: flow and charge transport,''
    Eur. Phys. J. C \textbf{80}, 293 (2020)
    % doi:10.1140/epjc/s10052-020-7847-4
    [arXiv:1908.07605 [hep-ph]].
    
%%%%%%%%%%%%%%%%%  Numerical (MDH) - Dynamic Conductivity 
 
    \bibitem{Yan:2021zjc}
    L.~Yan and X.~G.~Huang,
    %``Dynamical evolution of magnetic field in the pre-equilibrium quark-gluon plasma,''
    [arXiv:2104.00831 [nucl-th]].
    
    \bibitem{Wang:2021oqq}
    Z.~Wang, J.~Zhao, C.~Greiner, Z.~Xu and P.~Zhuang,
    %``Incomplete electromagnetic response of hot QCD matter,''
    [arXiv:2110.14302 [hep-ph]].
    
%%%%%%%%%%%%%%%%%%%%%%%%%%%%%%%%%%%%%%%%%%%%%%%%%%%%%%%%%%%%%%%%%%%%%%%%%%%

    \bibitem{Formanek:2021blc}
    M.~Formanek, C.~Grayson, J.~Rafelski and B.~M\"uller,
    %``Current-conserving relativistic linear response for collisional plasmas,''
    Annals Phys. \textbf{434}, 168605 (2021)
    % doi:10.1016/j.aop.2021.168605
     [arXiv:2105.07897 [physics.plasm-ph]].
    
    \bibitem{Florkowski:2017olj}
    W.~Florkowski, M.~P.~Heller and M.~Spalinski,
    %``New theories of relativistic hydrodynamics in the LHC era,''
    Rept. Prog. Phys. \textbf{81}, 046001 (2018)
    % doi:10.1088/1361-6633/aaa091
    [arXiv:1707.02282 [hep-ph]].
    
    \bibitem{Rocha:2021zcw}
    G.~S.~Rocha, G.~S.~Denicol and J.~Noronha,
    %``Novel Relaxation Time Approximation to the Relativistic Boltzmann Equation,''
    Phys. Rev. Lett. \textbf{127}, 042301 (2021)
    % doi:10.1103/PhysRevLett.127.042301
    [arXiv:2103.07489 [nucl-th]].

    \bibitem{Bhatnagar:1954zz}
	P.~L.~Bhatnagar, E.~P.~Gross and M.~Krook,
	%``A Model for Collision Processes in Gases. 1. Small Amplitude Processes in Charged and Neutral One-Component Systems,''
	Phys. Rev. \textbf{94}, 511 (1954).
	%doi:10.1103/PhysRev.94.511
	
    \bibitem{Song:2007ux}
    H.~Song and U.~W.~Heinz,
    %``Causal viscous hydrodynamics in 2+1 dimensions for relativistic heavy-ion collisions,''
    Phys. Rev. C \textbf{77}, 064901 (2008)
    % doi:10.1103/PhysRevC.77.064901
    [arXiv:0712.3715 [nucl-th]].
    
    \bibitem{Starke:2014tfa}
	R.~Starke and G.~A.~H.~Schober,
	%\lq\lq Relativistic covariance of Ohm's law,\rq\rq
	Int. J. Mod. Phys. D \textbf{25}, 1640010 (2016)
	%doi:10.1142/S0218271816400101
	[arXiv:1409.3723 [math-ph]].
	
    \bibitem{Weldon:1982aq}
	H.~A.~Weldon,
	%\lq\lq Covariant Calculations at Finite Temperature: The Relativistic Plasma,\rq\rq
	Phys. Rev. D \textbf{26}, 1394 (1982).
	%doi:10.1103/PhysRevD.26.1394
	
	\bibitem{Melrose:2008}
	D.~Melrose,
	{\it Quantum Plasmadynamics: Unmagnetized Plasmas},
	Lect. Notes Phys. \textbf{735} 
	(Springer, New York, 2008).
	%doi:10.1007/978-0-387-73903-8 

    \bibitem{Romatschke:2015gic}
    P.~Romatschke,
    %``Retarded correlators in kinetic theory: branch cuts, poles and hydrodynamic onset transitions,''
    Eur. Phys. J. C \textbf{76}, 352 (2016)
    % doi:10.1140/epjc/s10052-016-4169-7
    [arXiv:1512.02641 [hep-th]].
    
   \bibitem{Kapusta:1992fm}
    J.~I.~Kapusta,
    %``Screening of static QED electric fields in hot QCD,''
    Phys. Rev. D \textbf{46}, 4749 (1992)
    % doi:10.1103/PhysRevD.46.4749
    
    \bibitem{Jackson:2019mop}
    G.~Jackson,
    %``Two-loop thermal spectral functions with general kinematics,''
    Phys. Rev. D \textbf{100}, 116019 (2019)
    %doi:10.1103/PhysRevD.100.116019
    [arXiv:1910.07552 [hep-ph]].
    
    \bibitem{Mrowczynski:1988xu}
    S.~Mr\'owczy\'nski,
    %``On the Transport Coefficients of a Quark Plasma,''
    Acta Phys. Polon. B \textbf{19}, 91 (1988).
	
    \bibitem{Rafelski:2002}
    J.~Letessier, and  J.~Rafelski,
    {\it Hadrons and Quark-Gluon Plasma} 
    % (Cambridge Monographs on Particle Physics, Nuclear Physics and Cosmology). 
   ( Cambridge University Press, 2002)
    %doi:10.1017/CBO9780511534997
    
    \bibitem{Drude:1900}
	P.~Drude,
	Ann.\ Phys.\ (Leipzig) \textbf{1}, 566 (1900).
	
% 	\bibitem{Aarts:2020dda}
%     G.~Aarts and A.~Nikolaev,
%     %``Electrical conductivity of the quark-gluon plasma: perspective from lattice QCD,''
%     Eur. Phys. J. A \textbf{57}, 118 (2021)
%     % doi:10.1140/epja/s10050-021-00436-5
%      [arXiv:2008.12326 [hep-lat]].
    
    % %\cite{Brandt:2012jc}
    % \bibitem{Brandt:2012jc}
    % B.~B.~Brandt, A.~Francis, H.~B.~Meyer and H.~Wittig,
    % %``Thermal Correlators in the \textbackslash{}rho\textbackslash{} channel of two-flavor QCD,''
    % JHEP \textbf{03} (2013), 100
    % % doi:10.1007/JHEP03(2013)100
    % % [arXiv:1212.4200 [hep-lat]].
    % %103 citations counted in INSPIRE as of 10 Jan 2022
    
    % %\cite{Amato:2013naa}
    % \bibitem{Amato:2013naa}
    % A.~Amato, G.~Aarts, C.~Allton, P.~Giudice, S.~Hands and J.~I.~Skullerud,
    % %``Electrical conductivity of the quark-gluon plasma across the deconfinement transition,''
    % Phys. Rev. Lett. \textbf{111}, 172001 (2013)
    % % doi:10.1103/PhysRevLett.111.172001
    %  [arXiv:1307.6763 [hep-lat]].
    % %182 citations counted in INSPIRE as of 10 Jan 2022
    
    % %\cite{Aarts:2014nba}
    % \bibitem{Aarts:2014nba}
    % G.~Aarts, C.~Allton, A.~Amato, P.~Giudice, S.~Hands and J.~I.~Skullerud,
    % %``Electrical conductivity and charge diffusion in thermal QCD from the lattice,''
    % JHEP \textbf{02}, 186 (2015)
    % % doi:10.1007/JHEP02(2015)186
    %  [arXiv:1412.6411 [hep-lat]].
    % %168 citations counted in INSPIRE as of 10 Jan 2022
    
    % %\cite{Brandt:2015aqk}
    % \bibitem{Brandt:2015aqk}
    % B.~B.~Brandt, A.~Francis, B.~J\"ager and H.~B.~Meyer,
    % %``Charge transport and vector meson dissociation across the thermal phase transition in lattice QCD with two light quark flavors,''
    % Phys. Rev. D \textbf{93}, 054510 (2016) 
    % % doi:10.1103/PhysRevD.93.054510
    % [arXiv:1512.07249 [hep-lat]].
    % %53 citations counted in INSPIRE as of 10 Jan 2022
    
    % %\cite{Astrakhantsev:2019zkr}
    % \bibitem{Astrakhantsev:2019zkr}
    % N.~Astrakhantsev, V.~V.~Braguta, M.~D'Elia, A.~Y.~Kotov, A.~A.~Nikolaev and F.~Sanfilippo,
    % %``Lattice study of the electromagnetic conductivity of the quark-gluon plasma in an external magnetic field,''
    % Phys. Rev. D \textbf{102}, 054516 (2020) 
    % % doi:10.1103/PhysRevD.102.054516
    %  [arXiv:1910.08516 [hep-lat]].
    % %32 citations counted in INSPIRE as of 10 Jan 2022
    
    % %\cite{Greif:2014oia}
    % \bibitem{Greif:2014oia}
    % M.~Greif, I.~Bouras, C.~Greiner and Z.~Xu,
    % %``Electric conductivity of the quark-gluon plasma investigated using a perturbative QCD based parton cascade,''
    % Phys. Rev. D \textbf{90}, 094014 (2014) 
    % % doi:10.1103/PhysRevD.90.094014
    % [arXiv:1408.7049 [nucl-th]].
    % %89 citations counted in INSPIRE as of 10 Jan 2022
    
    
    % \bibitem{Satow:2014lia}
% 	D.~Satow,
% 	%\lq\lq Nonlinear electromagnetic response in quark-gluon plasma,\rq\rq
% 	Phys. Rev. D \textbf{90}, 034018 (2014)
% 	%doi:10.1103/PhysRevD.90.034018
% 	[arXiv:1406.7032 [hep-ph]].
    
    \bibitem{Kharzeev:2007jp}
    D.~E.~Kharzeev, L.~D.~McLerran and H.~J.~Warringa,
    %``The Effects of topological charge change in heavy ion collisions: 'Event by event P and CP violation',''
    Nucl. Phys. A \textbf{803}, 227 (2008)
    % doi:10.1016/j.nuclphysa.2008.02.298
    [arXiv:0711.0950 [hep-ph]].
    
    \bibitem{Bass:2000ib}
    S.~A.~Bass and A.~Dumitru,
    %``Dynamics of hot bulk QCD matter: From the quark gluon plasma to hadronic freezeout,''
    Phys. Rev. C \textbf{61}, 064909 (2000)
    % doi:10.1103/PhysRevC.61.064909
    [arXiv:nucl-th/0001033 [nucl-th]].
    
    \bibitem{Muller:2018ibh}
    B.~M\"uller and A.~Sch\"afer,
    %``Chiral magnetic effect and an experimental bound on the late time magnetic field strength,''
    Phys. Rev. D \textbf{98}, 071902 (2018) 
    % doi:10.1103/PhysRevD.98.071902
    [arXiv:1806.10907 [hep-ph]].
    
    \bibitem{Letessier:1992xd}
    J.~Letessier, A.~Tounsi, U.~W.~Heinz, J.~Sollfrank and J.~Rafelski,
    %``Evidence for a high entropy phase in nuclear collisions,''
    Phys. Rev. Lett. \textbf{70}, 3530 (1993)
    % doi:10.1103/PhysRevLett.70.3530
    [arXiv:hep-ph/9711349 [hep-ph]].



%   %\cite{Bjorken:1982qr}
%     \bibitem{Bjorken:1982qr}
%     J.~D.~Bjorken,
%     %``Highly Relativistic Nucleus-Nucleus Collisions: The Central Rapidity Region,''
%     Phys. Rev. D \textbf{27} (1983), 140-151
%     doi:10.1103/PhysRevD.27.140
%     %3331 citations counted in INSPIRE as of 03 Feb 2022
    
    % %\cite{Song:2010mg}
    % \bibitem{Song:2010mg}
    % H.~Song, S.~A.~Bass, U.~Heinz, T.~Hirano and C.~Shen,
    % %``200 A GeV Au+Au collisions serve a nearly perfect quark-gluon liquid,''
    % Phys. Rev. Lett. \textbf{106} (2011), 192301
    % [erratum: Phys. Rev. Lett. \textbf{109} (2012), 139904]
    % doi:10.1103/PhysRevLett.106.192301
    % [arXiv:1011.2783 [nucl-th]].
    % %453 citations counted in INSPIRE as of 17 Feb 2022
    
    % %\cite{Stewart:2021mjz}
    % \bibitem{Stewart:2021mjz}
    % E.~Stewart and K.~Tuchin,
    % %``Continuous evolution of electromagnetic field in heavy-ion collisions,''
    % Nucl. Phys. A \textbf{1016}, 122308 (2021)
    % % doi:10.1016/j.nuclphysa.2021.122308
    % [arXiv:2106.09124 [nucl-th]].
    % %3 citations counted in INSPIRE as of 12 Jan 2022
    
    \bibitem{DeVries:1987atn}
    H.~De Vries, C.~W.~De Jager and C.~De Vries,
    %``Nuclear charge and magnetization density distribution parameters from elastic electron scattering,''
    Atom. Data Nucl. Data Tabl. \textbf{36}, 495 (1987)
    % doi:10.1016/0092-640X(87)90013-1
    
    \end{thebibliography}
\end{document}